\documentclass{article}
\usepackage{graphicx} 

\usepackage{geometry}
\usepackage{units} 
\usepackage{dsfont}
\usepackage{authblk} 
\usepackage{amsmath}
\usepackage{graphicx}
\usepackage[colorlinks=true, allcolors=blue]{hyperref}
\usepackage{hyperref}
\usepackage{amsfonts}
\usepackage{booktabs} 
\usepackage{caption}  
\usepackage{graphicx}
\usepackage{enumitem}
\usepackage[T1]{fontenc}    
\usepackage{csquotes}       
\usepackage{setspace}

\usepackage{algorithm}

\usepackage{algpseudocode}
\usepackage{appendix}
\usepackage[square,numbers]{natbib}

\begin{document}
\renewcommand\thefootnote{\fnsymbol{footnote}}
\setcounter{footnote}{1}

\onehalfspacing

 \thispagestyle{empty}
\begin{center}

      \Large\textbf{Permutation Tests Based on the Copula-Graphic Estimator and Their Use for Survival Tree Construction}\\
      \vspace{1.5cm}
      \large{Pauline Baur \textsuperscript{1}
      \footnote{\textbf{Corresponding author:} Pauline Baur; \textbf{Email}: baur@statistik.tu-dortmund.de}, Markus Pauly \textsuperscript{1, 2}, Takeshi Emura \textsuperscript{3}}
      \vspace{1.5cm}
      
    \textsuperscript{1} Department of Statistics, TU Dortmund University, Dortmund, Germany\\
    \textsuperscript{2} Research Center Trustworthy Data Science and Security, UA Ruhr, Dortmund, Germany \\
    \textsuperscript{3} School of Informatics and Data Science, Hiroshima University, Hiroshima, Japan
    
    \vspace{2cm}

\end{center}

\noindent
\large{\textbf{Abstract}}
\vspace{0.7cm}

\noindent
Survival trees are popular alternatives to Cox or Aalen regression models that offer both modelling flexibility and graphical interpretability. This paper introduces a new algorithm for survival trees that relaxes the assumption of independent censoring. To this end, we use the copula-graphic estimator to estimate survival functions. This allows us to flexibly specify shape and strength of the dependence of survival and censoring times within survival trees. For splitting, we present a permutation test for the null hypothesis of equal survival. Our test statistic consists of the integrated absolute distance of the group’s copula-graphic estimators. A first simulation study shows a good type I error and power behavior of the new test. We thereby asses simulation settings of various group sizes, censoring percentages and grades of dependence generated by Clayton and Frank copulas. 
Using this test as splitting criterion, a second simulation study studies the performance of the resulting trees and compares it with that of the usual logrank-based tree.
Lastly, the tree algorithm is applied to real-world clinical trial data.
\vspace{1cm}


\renewcommand\thefootnote{}
\footnotetext{\textbf{Abbreviations:} CGE, copula-graphic estimator; PBC, Primary Biliary Cholangiti.}

\renewcommand\thefootnote{\fnsymbol{footnote}}
\setcounter{footnote}{1}

\newpage

\section{Introduction}

In survival analysis, censoring is a common phenomenon, occurring, for example, when participants are lost to follow-up in a clinical trial. Typically, both censoring times and survival times are random   \cite{leungCensoringIssuesSurvival1997}. Many common survival analysis methods are derived under the assumption of independent censoring and survival times   \cite{leungCensoringIssuesSurvival1997, kaplanNonparametricEstimationIncomplete1958}. However, this assumption 
may not always be realistic, potentially introducing bias to the survival analysis   \cite{huangRegressionSurvivalAnalysis2008, emuraGeneSelectionSurvival2016}. For example, Klein and Moeschberger (1987)  \cite{kleinIndependentDependentCompeting1987} explain this issue for the case of the Kaplan-Meier estimator. 

Dependent censoring can arise, for instance, when study dropouts occur due to adverse events or lack of improvement from the study medication   \cite{schneiderClaytonCopulaSurvival2023}, since a patient's health status affects both these events and the expected survival time. In case of a positive dependency between survival and censoring times, subjects censored at a time $t$ have a smaller expected survival compared to subjects censored at a time greater than $t$. Consequently, statistical methods assuming independent censoring can overestimate the survival time. The opposite is true for negative dependency   \cite{staplinDependentCensoringPiecewise2015}. This bias is especially problematic when two groups with varying censoring proportions are compared. One such scenario is a clinical placebo-controlled trial, where the verum group has a higher dropout rate due to study drug related adverse events. If the independent censoring assumption is violated,  applying biased survival time estimation can lead to overly optimistic survival estimates for the verum group  \cite{huangRegressionSurvivalAnalysis2008, templetonInformativeCensoringNeglected2020}. Hence, survival analysis methods that can model dependent censoring are often necessary.

Copulas are a common tool in such methods. They specify the joint distribution of random variables  \citep[Chapter 2]{nelsenIntroductionCopulas2006}, in our case event and censoring times. Furthermore, copula models can be used to analyze the bias introduced by false independent censoring assumptions, as Emura and Chen (2016)  \cite{emuraGeneSelectionSurvival2016} show for univariate feature selection using Cox-regression. Zheng and Klein (1995)  \cite{zhengEstimatesMarginalSurvival1995}  introduce a copula-based estimator for survival distributions, the copula-graphic estimator. It  can be considered an extension of the Kaplan-Meier estimator that incorporates dependence using a pre-specified copula model. Many applications of this estimator exist: Lo and Wilke (2010)  \cite{loCopulaModelDependent2010} extend the copula-graphic estimator for data with more than two competing risks using the class of Archimedian copulas. Huang and Zahng (2008)  \cite{huangRegressionSurvivalAnalysis2008} apply the work of Zheng and Klein to a regression setting. They model marginal competing risks  using Cox proportional hazard models, while the joint distribution is modeled on an assumed copula. Many further applications
of the copula-graphic estimator in survival regression settings have been explored, including frequentist approaches  \cite{yehSensitivityAnalysisSurvival2023, braekersACopula2005} as well as Bayesian methods  \cite{schneiderClaytonCopulaSurvival2023}.

The present paper uses the copula-graphic estimator to derive nonparametric survival trees that allow for dependent censoring. Traditional classification and regression trees, which both are nonlinear regression models, 
were introduced by Breiman et. al (1984)  \cite{breimanClassificationRegressionTrees1984a}. Various extensions of their work to survival analysis under the independent censoring assumption have been developed   \cite{bou-hamadReviewSurvivalTrees2011}. In the present paper, we extend the idea behind conditional inference trees  \cite{hothornUnbiasedRecursivePartitioning2006} and the work of Emura et al. (2023)  \cite{emuraSurvivalTreeBased2023} to the setting of dependent censoring. We will construct  trees of binary splits on single covariates using $p$-values of significance tests for survival difference as a splitting criterion. We will modify existing tree algorithms, such as 
logrank trees  \cite{ciampiStratificationStepwiseRegression1986}, by using a significance test that does not assume independent censoring. In doing so, we present a new survival analysis method that is straightforward to implement, easy to interpret and free from assumptions of parametric regression models. 

To this end, we propose a permutation test with the integrated, absolute distance of the copula-graphic estimators of two groups as a test statistic for the null hypothesis of equal survival distributions, assuming equal censoring distributions across groups. Permutation tests are a common tool in survival analysis  \cite{hothornUnbiasedRecursivePartitioning2006, doblerBootstrapPermutationbasedInference2018,ditzhausCASANOVAPermutationInference2023} and similar approaches to ours already exist: Pepe and Fleming (1989)  \cite{pepeWeightedKaplanMeierStatistics1989} introduce a class of Kaplan-Meier estimator based statistics. They extend their statistic by various weighting functions that reduce the impact of observations towards the end of a study when only few events are observed. Moradian et al. (2017)  \cite{moradianSplittingRulesSurvival2017} use a statistic of absolute distances of Kaplan-Meier estimators as a splitting criterion in a survival forest. They later extend their work using the copula-graphic estimator to create a survival forest that can account for dependent censoring  \cite{moradianSurvivalForestsData2019}. 

The rest of the paper will be structured as follows: Section \ref{methods} will review the copula-graphic estimator and survival trees and then propose a survival tree algorithm for dependently censored survival data. Section \ref{study1} and \ref{study2} will asses the algorithm's performance in two simulation studies. Lastly, in Section \ref{example} we will apply the survival tree algorithm to real-world data from the Mayo Clinic Primary Biliary Cholangitis clinical trial.

\section{Methods}\label{methods}
\subsection{Notation} \label{Notation}

We consider right-censored survival data for $n$ subjects. Event times are modelled by non-negative random variables $T_i$, corresponding censoring times by non-negative $C_i$ 
\begin{equation*}
    T_i \sim F, \quad C_i \sim G, \quad i = 1, \ldots n
\end{equation*}
with continuous, strictly increasing distribution functions $F$ and $G$, respectively. For each subject $i$, only $X_i = \min(T_i, C_i)$ and the censoring status $\Delta_i = \mathds{1}(X_i = T_i)$ can be observed with $\mathds{1}()$ being the indicator function. The $X_i$s are assumed to be independent, identically distributed random variables. In addition, we observe $p$ covariates, which are a realization of the random vector $\boldsymbol{Z}_i = (Z_{i1}, \dots, Z_{ip})^\top$. Thus, the observed dataset is $\{(x_i, \delta_i, \boldsymbol{z}_i); i = 1, \ldots, n\}$  with lowercase letters denoting the realizations of the respective random variables. Throughout, vectors are denoted using bold font. The probability of subject $i$ surviving past a time $t>0$ is given by the survival function 
    $S_T(t) = Pr(T_i > t) = 1-F(t)$,
which is continuous and strictly decreasing \citep[Chapter 2]{kleinBasicQuantitiesModels2003}. The corresponding censoring function is defined analogously as $S_C(t) = Pr(C_i > t)=1-G(t)$. Here, $T$ and $C$ denote independent copies of $T_i$ and $C_i$, respectively.

\subsection{Copula-Graphic Estimator}

In the following, $T$ and $C$ are not independent, instead their dependency will be modeled using bivariate copulas. These will be defined based on survival functions, rather than cumulative distribution functions, to fit our survival setting. The general properties of copulas remain valid \citep[Chapter 3]{emuraCopulaModelsDependent2018}. The joint survival function of $T$ and $C$ is 
\begin{equation*}
	Pr(T > t, C > s) =  \mathcal{C}\left(S_T(t), S_C(s)\right), \quad s,t > 0, 
\end{equation*}
with $\mathcal{C}$ being the copula. A feasible copula function $\mathcal{C}$ has to fulfill $\mathcal{C}(u,0) = \mathcal{C}(0,v)=0$, $\mathcal{C}(u,1)=u$ and $\mathcal{C}(1,v) = v$  for every $u,v \in [0,1]$. Furthermore, we require $\mathcal{C}$ to yield a probability mass on every rectangle in $[0,1]^2$, by  ensuring $\mathcal{C}(u_2, v_2) - \mathcal{C}(u_2, v_1) - \mathcal{C}(u_1, v_2) + \mathcal{C}(u_1, v_1) \geq 0$ for all $u_1, u_2, v_1, v_2 \in [0,1]$ with $u_1 \leq u_2$ and $v_1 \leq v_2$. \citep[Chapter 2]{nelsenIntroductionCopulas2006}.

The scope of this paper will be restricted to the class of Archimedian copulas, which have a closed-form expression and thus are convenient to work with. Archimedian copulas are generated by a function $\varphi$ via
\begin{equation} \label{archimedianCopula}
	\mathcal{C}\left( S_T(t), S_C(s)\right)	 = \varphi^{-1}	
	\left(
	\varphi \left(S_T(t)\right) + \varphi \left(S_C(s)\right)
	\right)	; \text{\citep[Chapter 3]{emuraCopulaModelsDependent2018}}
\end{equation}
for $\varphi: [0,1] \to [0, \infty]$ being continuous and strictly decreasing to $\varphi(1) = 0$ with pseudo-inverse
\begin{equation*}
    \varphi^{-1}(u) 
        = \begin{cases}
            \varphi^{-1}(u), \quad & 0 \leq u \leq \varphi(0) \\
            0, \quad & \varphi(0) \leq u \leq \infty.
        \end{cases}
\end{equation*}
Equation \eqref{archimedianCopula} defines a copula, if $\varphi$ is convex. One example for an Archimedian copula is the Clayton copula, which is generated by $\varphi(u) = \nicefrac{\left( u^{-\theta} - 1\right)}{\theta}$ for a parameter $\theta \in [-1, \infty ) \setminus \{0\}$ \citep[Chapter 4]{nelsenIntroductionCopulas2006}. 

Copulas provide an easy-to-interpret way of modelling the dependence structure, specifically the concordance of   $T$ and $C$. On a realization level, a concordant pair of observations $(t_1, c_1)$ and $(t_2, c_2)$ fulfills  $(t_1-t_2)(c_1-c_2) >0$; a discordant one $(t_1-t_2)(c_1-c_2) <0$. On  the population level, the scale-invariant Kendall's $\tau$ measures association between i.i.d. vectors $(T_1, C_1)$ and $(T_2, C_2)$ with
\begin{equation*}
    \tau = Pr\left(
                (T_1 - T_2)(C_1 - C_2) > 0 
            \right)
            - 
            Pr\left(
                (T_1 - T_2)(C_1 - C_2) < 0 
            \right),
\end{equation*}
which is the difference of the probability of concordance and the probability of discordance. For random variables $S_T(T)$ and $S_C(C)$ with dependence structure $\mathcal{C}$, Kendall's $\tau$ can alternatively be calculated as
\begin{equation}\label{CopulaTau}
     \tau = 4 \mathbb{E}\left(\mathcal{C}\left(S_T(T),S_C(C)\right)\right) -1 \overset{(\ast)}{=} 1 + 4\int\limits_0^1 \frac{\varphi(t)}{\varphi^\prime(t)} dt
\end{equation}
with the last equation  $(\ast)$ holding true for Archimedian $\mathcal{C}$ \citep[Chapter 5]{nelsenIntroductionCopulas2006}. Thus, the association depends on the copula, but not the respective marginal distributions. For the Clayton copula, Equation \eqref{CopulaTau} simplifies to $\tau = \nicefrac{\theta}{(\theta + 2)}$, yielding a straightforward way to specify the level of concordance \citep[Chapter 5]{nelsenIntroductionCopulas2006}. Consequently, the Clayton copula's limiting case of $\theta \to 0$ is the independence copula, which models independent event and censoring times as $\mathcal{C}\left( S_T(t), S_C(s)\right) = Pr(T > t) Pr(C > s) $. It can be generated from Equation \eqref{archimedianCopula} by choosing $\varphi(u) = -\log(u)$ \citep[Chapter 4]{nelsenIntroductionCopulas2006}. 

As explained in the introduction, the well-known Kaplan-Meier estimator does not consider a dependency of survival and censoring times. For scenarios, where the independent censoring assumption is not realistic, Zheng and Klein (1995) \cite{zhengEstimatesMarginalSurvival1995} introduced an alternative estimator, the copula-graphic estimator (CGE), which estimates the survival function under a known dependence structure described by a copula. 
 Rivest and Wells (2001) \cite{rivestMartingaleApproachCopulaGraphic2001} extend this work by deriving a closed-form expression of the CGE for survival and censoring functions under the assumption of an Archimedian copula with twice-differentiable generator $\varphi$. They start by requesting the naive estimate for the survival function $\hat{\pi}(t) = \nicefrac{1}{n}\sum_{i=1}^n \mathds{1}(X_i > t)$ at time $t>0$ to be equal to the Archimedian copula structure from Equation (\ref{archimedianCopula}) based on estimators of $S_T$ and $S_C$, rather than the theoretical survival and censoring function. This can be denoted as
\begin{equation*}
	\varphi^{-1}
		\left[
		\varphi(\hat{S}(X_i)) + \varphi(\hat{C}(X_i))
		\right]	
	= \hat{\pi}(X_i),
	\quad i \in \{1, \ldots, n\}
\end{equation*}
with $\hat{S}$ and $\hat{C}$ being the  CGEs of the survival and censoring function, respectively. Solving this equation for $\hat{S}$ and considering $n$ observations, yields the  CGE of the survival function:
\begin{equation}\label{cgeFormel}
	\hat{S}(t) 
		= \varphi^{-1} 
		\left[
			- \sum\limits_{X_i \leq t, \delta_i = 1} 
			\varphi(\hat{\pi}(X_i)) - \varphi\left(\hat{\pi}(X_i) - \frac{1}{n}\right)
		\right], \quad 0 \leq t \leq \max(X_i).
\end{equation} 
It is  a right-continuous, decreasing step-function. $\hat{S}(0)$ equals  1. Subsequently, there are negative jumps at  each $x_i$  associated with an event ($\delta_i = 1$). Using the generator of the independence copula $\varphi(u) = -\log(u)$  in Equation \eqref{cgeFormel}, the  CGE reduces to the Kaplan-Meier estimator \cite{rivestMartingaleApproachCopulaGraphic2001}. Furthermore, $\hat{S}(t)$ equals the Kaplan-Meier curve for  any $t$ greater than the last timepoint observed in a study \cite{zhengEstimatesMarginalSurvival1995}. A visualization of the  CGE and the influence of the assumed dependency through the copula model can be found in Figure \ref{clayton2} on page \pageref{clayton2} in the Appendix.

\subsection{Proposed Test}\label{sec:proposedTest}

In the following paragraphs, we propose a permutation test assessing differences in survival times between two groups using a randomization technique for the exact control of the type I error rate. The notation introduced in Section \ref{Notation} will be extended to cover a two-sample problem. For group $j \in \{1, 2\}$, the event and censoring time variables will be
\begin{equation*}
    T_{ji} \sim F_j, \quad C_{ji} \sim G_j, \quad  j = 1,2, \quad i = 1, \ldots n_j,
\end{equation*}
with respective distributions $F_j$ and $G_j$. Censored data $X_{ji}=\min(T_{ij},C_{ij})$ and $\Delta_{ji}=\mathds{1}(X_{ij} = T_{ij})$ are adapted accordingly. 
The null and alternative hypotheses for a difference in survival distributions of the two samples are given by
\begin{equation}\label{hypothesis}
 H_0: S_{T_{11}} = S_{T_{21}} \, \text{ vs. }  H_1: S_{T_{11}} \neq S_{T_{21}}.
\end{equation}
The censoring distributions for both groups, $G_1$ and $G_2$, are assumed to be identical, such that the $X_{ij}$ are exchangeable under the null hypothesis.
We will use the CGE for Archimedian copulas to construct a test statistic for a non-parametric permutation test. The test statistic is similar to the one proposed by Moradian et al. (2019)  \cite{moradianSurvivalForestsData2019}, who used a similar statistic as a measure of prognostic difference between groups for determining optimal splits in a survival random forest. We slightly modify the statistic and will introduce a permutation test and corresponding splitting criterion.\newline
Intuitively, the statistic is derived from the absolute difference of the  CGEs of group 1 and 2 and should increase with differing $ S_{T_{11}}$ and $S_{T_{21}}$. For observation vector  $\boldsymbol{x} = (\boldsymbol{x_1}^\top, \boldsymbol{x_2}^\top)^\top$ with observations in group $j$ being  $\boldsymbol{x_j} = \left(\min(t_{j1}, c_{j1}), \ldots, \min(t_{jn_j}, c_{jn_j})\right)^\top$ and censoring indicator vector  $\boldsymbol{\delta} = (\boldsymbol{\delta_1}^\top, \boldsymbol{\delta_2}^\top)^\top$, the statistic is
\begin{equation}\label{eq:L1stat}
 L_1(\boldsymbol{x}, \boldsymbol{\delta}) = \int\limits_{\min(\boldsymbol{x_{1}},\boldsymbol{x_{2}})}^{\min(\max(\boldsymbol{x_{1}}),\max(\boldsymbol{x_{2}}))}  \frac{|\hat{S}_{T_{11}}(t) - \hat{S}_{T_{21}}(t)|}{\min(\max(\boldsymbol{x_{1}}),\max(\boldsymbol{x_{2}}))} dt.
\end{equation}
The division by $\min(\max(\boldsymbol{x_{1}}),\max(\boldsymbol{x_{2}}))$ accounts for the observation spans and is a modification compared to the statistic used by Moradian et al. (2019) \cite{moradianSurvivalForestsData2019}. Since the theoretical probability distribution of $L_1$ is difficult to be derived analytically, we will evaluate it using a randomization approach. 

To find the permutation distribution of $L_1$ and calculate corresponding $p$-values, we consider the finite permutation group  
\begin{equation*}
    \mathcal{G} = \left\{g: \mathbb{R}^{2\times n} \to \mathbb{R}^{2\times n}, 
        \left(\begin{pmatrix}x_1\\ \delta_1 \end{pmatrix}, \ldots, \begin{pmatrix}x_n\\ \delta_n \end{pmatrix} \right) 
    \mapsto 
        \left(\begin{pmatrix}x_{\pi(1)}\\ \delta_{\pi(1)} \end{pmatrix}, \ldots, \begin{pmatrix}x_{\pi(n)}\\ \delta_{\pi(n)} \end{pmatrix} \right) 
        \right\}
\end{equation*} of size  $\vert\mathcal{G}\vert = n!$ for any permuting function $\pi: \{1, \ldots, n\} \to \{1, \ldots, n\}$. Enabled by the assumption of exchangeability  and equal censoring distributions across groups,  the distribution of $(\boldsymbol{X}, \boldsymbol{\Delta})$ with  $\boldsymbol{X} = \left( X_{11}, \ldots, X_{1n_1}, X_{21},\ldots, X_{1n_2}\right)$ is invariant to permutations in $\mathcal{G}$ under the null hypothesis. In this case, the following test $\psi$ based on statistic $L_1$ is an exact level $\alpha$ test for $H_0$ as in Equation \eqref{hypothesis}, i.e. $\mathbf{E}\left(\psi(\boldsymbol{X}, \boldsymbol{\Delta})\right) = \alpha$  \citep[Chapter 17]{lehmannPermutationRandomizationTests2022}:
\begin{equation} \label{RandomizationTest}
    \psi(\boldsymbol{x}, \boldsymbol{\delta}) =
        \begin{cases}
            1,                  & \text{ if }  L_1(\boldsymbol{x}, \boldsymbol{\delta}) > L_1^{(k)}(\boldsymbol{x}, \boldsymbol{\delta})   \\
            a(\boldsymbol{x}, \boldsymbol{\delta}),  & \text{ if }  L_1(\boldsymbol{x}, \boldsymbol{\delta}) = L_1^{(k)}(\boldsymbol{x}, \boldsymbol{\delta})   \\
            0,                  & \text{ if }  L_1(\boldsymbol{x}, \boldsymbol{\delta}) < L_1^{(k)}(\boldsymbol{x}, \boldsymbol{\delta}).
        \end{cases}     
\end{equation}
$ L_1(\boldsymbol{x}, \boldsymbol{\delta})$ is the test statistic on the observed data and critical value $L_1^{(k)}(\boldsymbol{x}, \boldsymbol{\delta})$ is derived by calculating the test statistics for all $\vert\mathcal{G}\vert$ permutations, ordering them to  $L_1^{(1)}(\boldsymbol{x}, \boldsymbol{\delta}) \leq L_1^{(2)}(\boldsymbol{x}, \boldsymbol{\delta}) \leq \cdots \leq L_1^{(n!)}(\boldsymbol{x}, \boldsymbol{\delta})$ and considering the  $k = (n! - \lfloor n! \alpha \rfloor)$th value. By setting randomization probability
 $a(\boldsymbol{x}, \boldsymbol{\delta}) = \left(\alpha n! - \vert\{j: L_1^{(j)}(\boldsymbol{x}, \boldsymbol{\delta}) > L_1^{(k)}(\boldsymbol{x}, \boldsymbol{\delta}) \}\vert\right) \cdot \left(\vert \{j: L_1^{(j)}(\boldsymbol{x}, \boldsymbol{\delta}) = L_1^{(k)}(\boldsymbol{x}, \boldsymbol{\delta}) \} \vert \right)^{-1}, j \in \{1, \ldots n!\}$, we ensure that $\psi$ is an exact level-$\alpha$-test. 
 
In most cases, a systematic calculation of all $n!$ permutations would be intangible and we therefore resort to $n_{perm}\leq n!$ random data permutations. 
The resulting algorithm for its $p$-value computation is given below:  
\begin{algorithm}
\caption{$p$-value of introduced randomization test}\label{alg:pvalue}
\begin{algorithmic}
\State Calculate statistic  $L_{1obs} = L_1(\boldsymbol{x}, \boldsymbol{\delta})$ on observed data
\For{$i \in \{1, \ldots n_{perm}\}$} 
\State Randomly permute $(\boldsymbol{x}, \boldsymbol{\delta})$ to get $(\boldsymbol{x}_{\pi(i)}, \boldsymbol{\delta}_{\pi(i)})$
\State Calculate statistic  $L_{1perm_i} = L_1(\boldsymbol{x}_{\pi(i)}, \boldsymbol{\delta}_{\pi(i)})$. 
\EndFor 
\State Calculate $p$-value from statistics $ L_{1perm_i}$:
\begin{equation*}
	 p_{perm} = \frac{\sum\limits_{i=1}^{n_{perm}} \mathds{1}\{ L_{1perm_i} \geq L_{1obs} \}+ 1}{n_{perm} + 1}
\end{equation*}
\end{algorithmic}
\end{algorithm}

\subsection{Survival Trees} \label{sec:survival.trees}

We will construct survival trees using  recursive partitioning, which repeatedly splits the covariable space into two disjoint sub-spaces,  yielding increasingly  homogeneous survival outcomes within and heterogeneous outcomes between groups. A basic method of tree building uses binary splits based on a single covariate at a time \cite{bou-hamadReviewSurvivalTrees2011}. Each node partitions the data into  child nodes $\{i: z_{ij} \leq q\}$ and $\{i: z_{ij} > q\}$ based on the $j$th covariable and some cutoff value $q$, for $i \in \{1, \ldots, n\}$ and $j \in \{1, \ldots, p\}$. 

Each split is chosen to maximize the prognostic survival difference between the resulting groups. Ciampi et al. (1986)  \cite{ciampiStratificationStepwiseRegression1986}  use a logrank test as a measure for prognostic difference. Performing a grid search over all covariables and feasible cutoff values, logrank test $p$-values are calculated for all possible splits. The optimal split is chosen as the split that maximizes the test statistic among all significant tests to some pre-defined $p$-value threshold. Nodes are split until no feasible split with a $p$-value smaller than the threshold can be found. Emura et al. (2023) \cite{emuraSurvivalTreeBased2023} generally adapt this approach, but select the split that minimizes the $p$-value. We will follow their approach, but use the test introduced in Section \ref{sec:proposedTest} instead of a logrank test, in order to account for possibly dependent censoring. Moradian et al (2019)  \cite{moradianSurvivalForestsData2019}
 already used a similar statistic to construct survival trees, however, their statsitic was not evaluatet within a statistical test and they constructed a random forest rather than a single tree.
 
Typically, median survival time and Kaplan-Meier estimator of the resulting terminal nodes are reported \cite{bou-hamadReviewSurvivalTrees2011}. We will supplement this by the  CGE. For ideal interpretation of the tree, we aim to order the terminal nodes by survival prognosis from left to right.  To do so, the test statistic in \eqref{eq:L1stat} is calculated without absolute values, denoted as $\tilde{L}_1$, such that positive values hint a longer survival in group 1. Depending on the sign of $\tilde{L}_1$, group one is detected to the left or right child of a node. This yields the following tree algorithm:

\begin{algorithm}[h]
\caption{Construction of survival tree}\label{alg:tree}
\begin{algorithmic}
\State \textbf{Step 0:}  Choose threshold $\tilde{p} \in [0,1]$.
\While{$n>2$ holds for current node} 
    \For{$j \in \{1, \ldots p\}$ and $k \in \{q_k$ feasible cutoff value$ \}$} 
          \State Calculate $p$-value $p_{jk}$ for  $H_0: S_{T}(t|  z_{ij} \leq q_k) = S_{T}(t|  z_{ij} > q_k) $ using Algorithm \ref{alg:pvalue}
    \EndFor
    \State  Set $(j^*, k^*) = \arg\min_{j,k}\{p_{jk}\}$
    \If{$p_{j^*k^*} < \tilde{p}$}
        \State Calculate $\tilde{L}_1$
        \If{$\tilde{L}_1 \leq 0$}
             \State Assign left child: $\{i: z_{ij^*} \leq q_{k^*}\}$; right child: $\{i: z_{ij^*} > q_{k^*}\}$
        \Else    
           \State     Assign left child: $\{i: z_{ij^*} > q_{k^*}\}$; right child: $\{i: z_{ij^*} \leq q_{k^*}\}$
        \EndIf   
    \Else
        \State Current node is terminal node
    \EndIf
\EndWhile
\end{algorithmic}
\end{algorithm}

\subsection{Choice of Copula Model and Dependency Parameter}

Since the true marginal distribution of survival and censoring times is not identifiable from an observed competing risk dataset \cite{tsiatisNonidentifiabilityAspectProblem1975},  additional assumptions have to be made prior to estimation. One assumption providing identifiability is the copula assumption  \cite{zhengEstimatesMarginalSurvival1995} introduced above.  Therefore selecting a sensible copula generator $\varphi$ and dependence $\tau$ in Equation (\ref{cgeFormel}) will be crucial and challenging step for our data analysis.

Zheng and Klein (1995) \cite{zhengEstimatesMarginalSurvival1995} evaluate the robustness of the CGE under a misspecified copula model. They find that, as long as the strength of dependency between survival and censoring times is estimated well, the CGE is relatively robust towards misspecification of the copula class. 
Therefore, we will decide on one copula class, namely the Clayton copula, prior to our data analysis. 
The Clayton copula can model  positive and negative dependency and specifies a straightforward and easy to interpret connection between it's parameter $\theta$ and Kendall's $\tau$  \citep[Chapter 4,5]{nelsenIntroductionCopulas2006}. 
Furthermore, the Clayton copula has successfully been used in previous research to model dependency in survival times, for instance showing good results in modelling the time to metamorphosis for salamander larvae \cite{emuraCopulabasedInferencePiecewise2017}, or in analyzing adherence to tuberculosis treatment  \cite{schneiderClaytonCopulaSurvival2023}.
Lastly,  due to the Clayton copula's simple generator function, we were able to implement a completely vector and matrix-based algorithm version of the permutation test in Algorithm \ref{alg:pvalue} in \texttt{R}, which saved computing time.

While there are approaches of estimating the dependency parameter from the data, these approaches typically lead to estimators with large variances, especially for small sample sizes  \cite{hsuReliabilityInferenceCopulaBased2016}. To avoid this,  a sensitivity analysis can be applied by varying the dependence parameter, evaluating model performance and then using the results to draw conclusion on the underlying dependence structure  \cite{staplinDependentCensoringPiecewise2015}. Emura and Chen (2016) \cite{emuraGeneSelectionSurvival2016}, who introduce an extension of univariate Cox regression for dependent censoring, recommend selecting $\tau$ using a cross-validated Harrell's $C$-index. They build their model for various assumed $\tau$, estimate Harrell's $C$ for each model and then choose the model with the largest $C$-index. We will use the same approach for our model selection and add the Integrated Brier Score as an additional performance measure. More details can be found in Section \ref{sec:tree.study.design}, where the setup of our survival tree study is explained.

\section{Simulation Study: Tests}\label{study1}
\subsection{Simulation Design}\label{sec:simuation.design.test}

The following simulation study aims to show that the proposed test indeed is a level-$\alpha$ test with good power properties. Furthermore, we  identify scenarios where the proposed test might fail. The simulation study is inspired by studies assessing properties of classification models  \cite{emuraGeneSelectionSurvival2016, emuraSurvivalTreeBased2023, moradianSurvivalForestsData2019}, in a sense that it provides simulation settings that can be generalized to datasets  described in Section \ref{Notation}.

Survival times are simulated assuming a survival function from a Cox proportional hazard model \citep[Chapter 2]{kleinBasicQuantitiesModels2003}. We condition on a one-dimensional observed covariate $z$ and receive $S_T(t|z) =\exp\left( -H_0(t) \exp(\beta z) \right)$ for a parameter $\beta\in \mathbb{R}$ and cumulative baseline hazard function $H_0$. Since  $\exp\left( -H_0(T) \exp(\beta z) \right) \sim \text{Unif}[0,1]$  for survival time $T$, we solve this term for $T$, generate unifrom random variables $U$ and then simulate our survival times as $T=H_0^{-1}\left(-\log(U)\exp(-\beta z) \right)$ \cite{benderGeneratingSurvivalTimes2005}. Censoring times are generated by drawing a second set of times, similar to the one described before, and setting the minimum of event and censoring time as the observed value \cite{crowtherSimulatingBiologicallyPlausible2013}.  

Since previous simulation studies on survival data showed censoring rates to influence results much more than underlying statistical distributions, we choose simple distributions and focus on modeling our tests properties under various censoring rates  \cite{dormuthComparativeStudyAlternatives2023}. To achieve this, we generate survival times using an exponential model with scale parameter $\lambda =1$, such that $H_0(t) = \lambda t$ and $T = - \log(U)\exp(-\beta z)$. Censoring times are generated independently from covariates and $\beta$, to ensure equal censoring distributions across groups throughout the whole simulation study. The scale parameter is set to  $\lambda = \left( \nicefrac{1}{(1-r)} \right) - r$, such that for $\beta = 0$ and independent survival and censoring times, the resulting survival data will have a censoring percentage of $100r$ \% \cite{wanSimulatingSurvivalData2017}. This leads to censoring time $C = -\log(V)\cdot \left( \left(  \nicefrac{1}{(1-r)} \right) - r \right)^{-1}$ for uniform $V$.

To model dependent survival times, $U,V \sim \text{Unif}[0,1]$ are drawn according to a pre-specified copula model using the \texttt{R}-package \texttt{Copula.surv} \cite{emuraCopulaSurvAnalysis2022}.

For the major part of the study,  we consider binary covariates with $z_{i}=0$ for subjects from group 1 and $z_{i}=1$ for subjects from group 2. Parameter $\beta$ varies on $[-1.4, 1.4]$, with $\beta=0$ representing the null hypothesis. Tryout simulations indicated that the interval of $[-1.4, 1.4]$ is large enough to observe a power close to 1 towards the edge of the considered $\beta$-interval. For type I error analysis, we consider all group sample sizes in $(n_1, n_2) \in \{20, 50, 100, 200\} \times \{20, 50, 100, 200\}$.  The power analysis considers sample sizes in $(n_1, n_2) \in \{20, 50, 150\} \times \{20, 50, 150\}$. Three stages of censoring are simulated by choosing $r \in \{0.1, 0.25, 0.5\}$. All simulation settings are tested on  two copula models, namely the Frank copula and the Clayton copula, with true theoretical dependency parameters $\tau_{theor.} \in \{0.0001, 0.25, 0.5, 0.75\}$. We set the desired type I error to $\alpha=0.05$.

Besides the described  regression-inspired designs with survival times depending on some $\beta z$, we  additionally assess power on alternatives with varying covariate generating mechanisms between groups and $\beta = 1$  for both groups, to mimic datasets like the one we will study in Section \ref{sec:data_example} with survival times possibly depending on various clinical covariates. Namely, we consider:

\textbf{1.} Normally distributed covariates with mean $\mu=0$ in group 1, $\mu=\gamma$ in group 2 and standard deviation $\sigma=1$ in both groups. Parameter $\gamma$ is varied on $[-1.5, 1.5]$ with $\gamma =0$ describing the null hypothesis.

\textbf{2.}  Normally distributed covariates with variance $\sigma=1$ in group 1, $\sigma\gamma$ in group 2 and mean $\mu=0$ in both groups. Parameter $\gamma$ is varied on $[0.00001, 10]$ with $\gamma =1$ describing the null hypothesis.

\textbf{3.}  Poisson distributed covariates with  parameter $\lambda=1$ in group 1 and parameter $\lambda=1+\gamma$ in group 2. Parameter $\gamma$ is varied on $[-0.9,1.5]$ with $\gamma =0$ describing the null hypothesis.

The  type I error, or power, from $n_{sim}$ rounds of simulations is estimated as $\widehat{\text{power}}= \nicefrac{1}{n_{sim}} \sum_{i=1}^{n_{sim}}$ $\mathds{1}(p_{perm_i} \leq \alpha)$ for $p_{perm_i}$ being the $p$-value from the $i$th simulation. The uncertainty of performing a simulation study with a finite number of repetitions  is given by the Monte-Carlo standard error (SE) as  $\sqrt{\nicefrac{1}{n_{sim}}\left(\widehat{\text{power}} \times (1-\widehat{\text{power}} )\right)}$ \cite{morrisUsingSimulationStudies2019}. Thus, for an estimated  type I error of $0.05$ from  $n_{sim}=2000$ repetitions, the Monte-Carlo standard error estimate would be 0.005.  During power estimation its upper bound is given by $\nicefrac{0.5}{\sqrt{n_{sim}}}$ \cite{marozziMultivariateTestsBased2016}, which is 0.011 for $n_{sim}=2000$ and  0.016 for $n_{sim}=1000$. 

The additional insecurity of calculating a permutation test on $n_{perm}$ rather than all $n!$ permutations can approximately be quantified by a factor of 1.2 that is added to the Monte-Carlo standard error \cite{boosMonteCarloEvaluation2000}. Both for the type I error and maximal error based on 1000 or 2000 permutations, this seems acceptable for the purpose of getting a general idea of our test's performance. Thus, we choose  $n_{sim}=1000$ for calculating power curves and $n_{sim}=2000$ for type I error analysis, where the exact maintenance of the error rate seems relevant. A permutation number of $n_{perm}$ = 1000 is chosen for both cases, which is well above the suggestion of $8\sqrt{n_{sim}}$  by Boos and Zhang (2000) \cite{boosMonteCarloEvaluation2000}.

We compare  the performance of the permutation  test  introduced in Section \ref{sec:proposedTest} using the Clayton copula and assumed concordance parameter $\theta_{assum.} \in \{ 0.000, 0.\bar{6}, 2, 6 \}$, which corresponds to  $\tau_{assum.} \in \{0.000, 0.25, 0.5, 0.75\}$ \footnote{The first entry of $\theta_{assum.}$ is in fact 0.00020002 and that of $\tau_{assum.}$ is  0.0001. They were chosen slightly larger than 0 to use the same algorithms as for larger $\tau$.}. Since logrank test-based survival trees are commonly found in literature  \cite{bou-hamadReviewSurvivalTrees2011, ciampiStratificationStepwiseRegression1986},  we additionally included  the logrank test, calculated with the \texttt{R}-package \texttt{survival} into our simulation study \cite{survival-package}.

The software \texttt{R} in version 4.2.1 was used for all calculations  \cite{rcoreteamLanguageEnvironmentStatistical2022} and visualizations were made using the \texttt{ggplot2} package \cite{wickhamGgplot2ElegantGraphics2016}.

\subsection{Simulation Results}\label{sec:test.study.result}

In the following, the results of type I error and power analysis will be shown. Unless stated otherwise, the described results refer to the settings with Clayton copula modelled dependence and binary covariates. Numbers will be rounded to three digits after the decimal sign.

All tests maintain the type I error rate of 0.05 relatively well in all settings, with 90\% of type I error estimates falling into the interval of $[0.043, 0.060 ]$ and all estimates being within $[0.037 , 0.0685]$. The median type I error estimate is 0.051. An overview of the estimated type I error rates for moderate sample sizes ($n_1=n_2=50$) on data generated by the Clayton copula, which in our case is the correctly specified copula class, can be seen in Table \ref{tab:type1n50} on page \pageref{tab:type1n50} in the Appendix. 
While the  CGE-based tests show acceptable type I error estimates in all cases, Table \ref{tab:type1n50} does not display any coherent patterns across varying $\tau_{theor.}$ and censoring scenarios. Similar findings were made for other sample sizes.

The logrank test's type I error does not increase substantially when its assumption of independent survival and censoring times is violated.  For small sample sizes ($n_1=20$ or $n_2=20$)  and a high dependency parameter of $\tau_{theor.}=0.75 $ all tests, but in particular the logrank test, are slightly  too liberal. The logrank test has type I error estimates up to 0.067 (SE of 0.006), which can be seen in Table \ref{tab:type1n20} in the Appendix. For larger $n$, this problem vanishes, as Table \ref{tab:type1n200}  illustrates.

We did not find any systematic difference of type I errors between tests on data generated using a Clayton copula and tests on Frank copula data, which result in a misspecified model. Both copula scenarios had a median type I error of 0.051 over all considered simulation settings, with 90\% of the type I errors falling into $[0.0435, 0.0588 ]$ for the Clayton copula and into  $[0.043, 0.060]$ for the Frank copula, respectively. Figure \ref{fig:type1error.by.copula} on page \pageref{fig:type1error.by.copula} gives more insight into this and compares type I error rates across copula models by test, sample size and theoretical dependence $\tau_{theor.}$.  The data displayed in the graphic is generated with censoring parameter $r=0.5$, yielding a mean censoring proportion of $0.500$. The considered misspecification of the copula class do not seam to cause inflated or too conservative type I error rates. Furthermore, no trend in type I error rates for varying $\tau_{theor.} $ is visible for data from either copula model. In particular, our study results do not show a superior maintenance of the type I error rates, when the true $\tau_{theor.}$ and the assumed $\tau_{assum.}$ coincide. Again, the logrank test is slightly more liberal than the  CGE-based test. The results for $r\in \{0.1, 0.25\}$ are similar.

Overall, we are satisfied with the type I error of the proposed test and move on to power analysis.

\begin{figure}[t]
\centering
\includegraphics[scale=0.54]{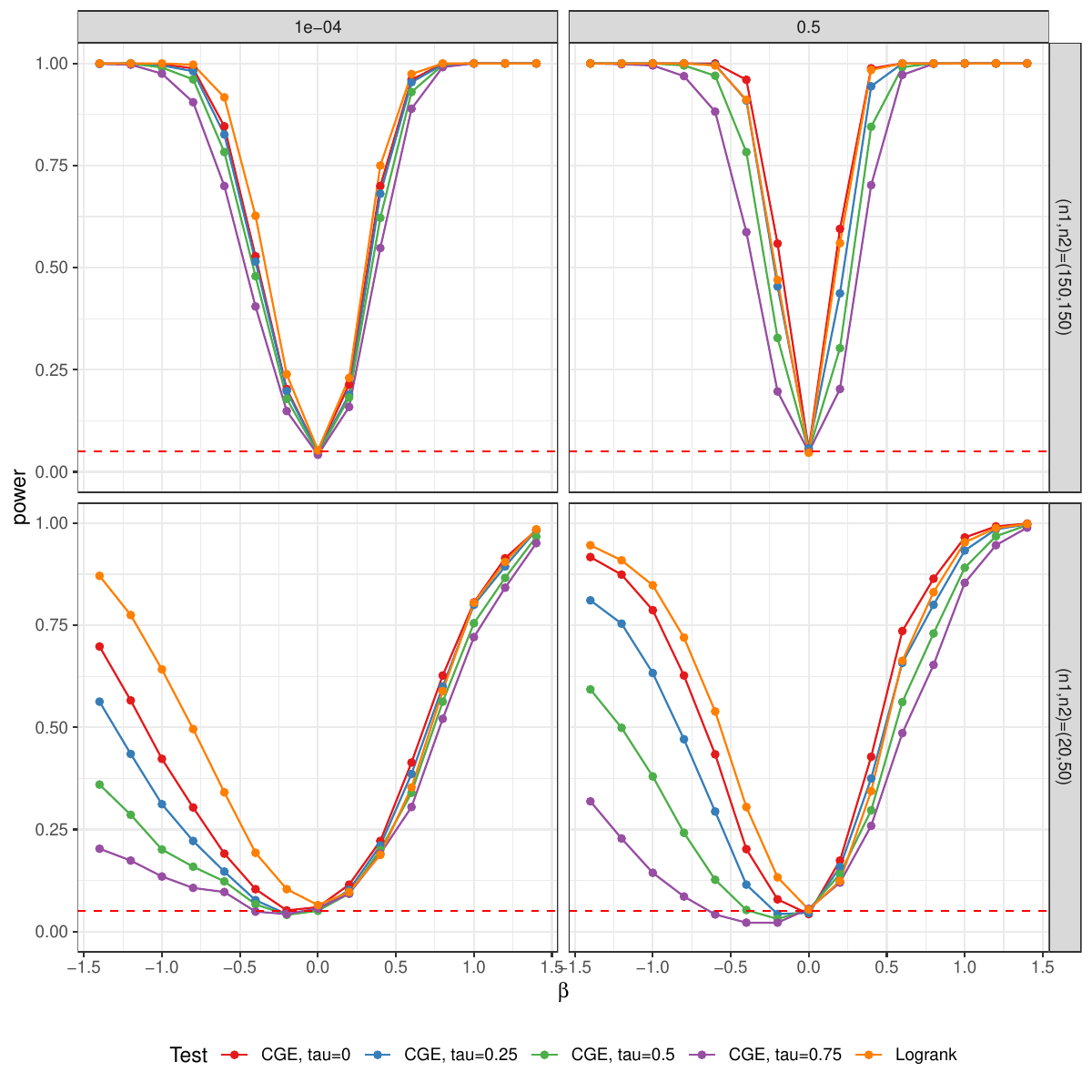}
\caption{Power estimates with theoretical dependency of event and censoring times of $\tau_{theor.} = 0.0001$ (left),  and  $\tau_{theor.} = 0.5$ (right) for  $n_1=n_2=150$ (top) and $n_1 = 20, n_2 = 50$ (bottom).  $r=0.5$ for all Figures. \label{plot:power.large.n}}
\end{figure}

The power estimates for large sample sizes of $n_1 = n_2 = 150$, binary covariates and  simulated dependency of $\tau_{theor.}\in \{0.0001, 0.5\}$ can be seen in the  upper part of  Figure \ref{plot:power.large.n}. The data for the graphic was generated with high censoring parameter $r=0.5$. Two things have to be noted: Firstly, the censoring distributions of both groups are identical and independent from $\beta$. However, the survival time distributions on the alternative hypothesis vary between groups. Our observations were generated as $\boldsymbol{x}= \min(\boldsymbol{t}, \boldsymbol{c})$ for survival times $\boldsymbol{t}$ and censoring times $\boldsymbol{c}$. Therefore, the mean censoring percentage of our simulated data in group 1 (which has survival times not affected by $\beta$) is constantly around 0.500, but the mean censoring percentage in group 2 varies with $\beta$.  The deviations in the censoring proportion in both groups range from minor ones (e.g. 0.501 vs.  0.465 for $\beta = 0.2$) to major ones (e.g. 0.500 vs. 0.337 for $\beta = 1$). More detailed information on censoring percentages of the data from Figure \ref{plot:power.large.n} can be found in Table \ref{tab:censoring.power}. Secondly, the censoring mechanism seems to not be completely independent from $\tau_{theor.}$. Censoring proporions rise with $\tau_{theor.}$ for negative $\beta$ (e.g. 0.591 for $\tau_{theor.}$ = 0.0001  and 0.646 for $\tau_{theor.}$ = 0.5 for $\beta = -0.6$ and $r=0.5$) and fall for positive $\tau_{theor.}$ (e.g. 0.425 for $\tau_{theor.}$ = 0.0001  vs. 0.351 for $\tau_{theor.}$ = 0.5 for $\beta = 0.6$). Thus, while a comparison of the tests within each sup blot in Figure \ref{plot:power.large.n} is possible, since the powers were estimated on the same datasets, only a limited interpretation of the performance for varying dependence is reasonable.

All in all, the tests were able to detect deviations from the null hypothesis $\beta=0$ and for large $|\beta|$ their power estimates are close to 1. In all  cases, the  CGE tests with lower $\tau_{assum.}$ performed better with the   $\tau_{assum.}=0$-test having the best power. 

Exemplary, for $\tau_{theor.} =0.5$ and $\beta = -0.4$, the  CGE-based test with $\tau_{assum.}=0$ has an estimated power of 0.960 (SE of 0.001). The test for  $\tau_{assum.}=0.5$ has a power of 0.783 (SE of 0.013) and the test with $\tau_{assum.}=0.75$ is at 0.587 (SE of  0.016). 

All tests showed a faster increase with positive $\beta$ than with negative $\beta$, which partially could be attributed to the varying mean censoring percentages described in Table \ref{tab:censoring.power}. In both displayed scenarios with $n_1 = n_2= 150$ of Figure \ref{plot:power.large.n}, the logrank test and the  CGE-based test for  $\tau_{assum.}=0$  perform similarly. While for $\tau_{theor.} \in \{0.0001, 0.25\}$, the logrank test has the higher power, the  CGE test  has a higher power for $\tau_{theor.} = 0.5$. For simulation settings with balanced, but smaller group sizes, the power rises slower than for the case of $n_1 = n_2 = 150$, but the performance of the test in relation to each other remains the same. See Figure \ref{plot:power.small.n} and Figure \ref{plot:power.medium.n} in the Appendix for details.

The lower part of Figure  \ref{plot:power.large.n} illustrates deficits  of the CGE tests in  unbalanced, small sample size settings.  
In the case of $n_1=20$ and $n_2 = 50$, the logrank test's power rises steadily with  $\beta<0$ increasing in absolute value. However, the  CGE-based test's power, especially for large  $\tau_{assum.}$ increases much slower and even falls to 0.030 (SE of 0.005) at $\beta=-0.4$ for the test with $\tau_{assum.}=0.75$. For $\beta<0$,  group 2 is expected to have longer survival times. This leads to scenarios, where the  latest observed time in  smaller group 1, $\max(\boldsymbol{x_{1}})$, is a lot lower than the latest time $\max(\boldsymbol{x_{2}})$ in group 2. An example of a single dataset from our simulation study illustrates this  in Figure \ref{plot:cge.ex.curve} in the Appendix. The logrank test is able to detect the longer survival of group 2 with a  $p$-value of  0.031. However, the  CGE-based statistic, only comparing survival curves up to $\min(\max(\boldsymbol{x_{1}}),\max(\boldsymbol{x_{2}}))$,  misses out on the fact that there is a large difference in $\max(\boldsymbol{x_{1}})$ and $\max(\boldsymbol{x_{2}})$ and yields a $p$-value of 0.170. Apparently, for this setting, the division of the absolute distance in between the  CGEs by observation time span $\min(\max(\boldsymbol{x_{1}}),\max(\boldsymbol{x_{2}}))$ is not enough to counteract this. The issue occurred for various values of $\tau_{theor.}$ (see Figure \ref{plot:power.unbalanced.bytau}), but was less present for lower censoring parameters $r$ (see Figure \ref{plot:power.unbalanced.byr}).

The problem vanishes, when group 1 has a larger sample size compared to group two, e.g. for $n_1=50$ and $n_2 = 20$. The larger sample size in group 1 with expected shorter survival leads to smaller differences of $\max(\boldsymbol{x_{1}})$ and $\max(\boldsymbol{x_{2}})$ and thus a larger proportion of the study time is accounted for in the CGE-based test statistic. In some settings, the  CGE tests outperform the logrank test, such as the case of $\beta=-0.2$, where the logrank test has a power of 0.057 (SE of 0.007)  and the four  CGE-based tests have powers inbetween 0.089 (SE of 0.009) and 0.098 (SE of 0.009). Again, an illustration of an exemplary dataset is shown in Figure \ref{plot:cge.ex.curve}.

All settings were evaluated on dependency generated by the Clayton and Frank copulas. No differences in test performance were seen.  Figure \ref{plot:power.copula} in the Appendix exemplarily illustrates this for one setting ($n_1=n_2=50$ and $r=0.25$).

In the following paragraphs, we will discuss results from the three settings with alternative covariate structures.
In setting \textbf{1.} additional variance is added to simulated times by adding normal covariates of varying means to group 2. Still, the means of the covariates generating the survival times are varied across the same range as $\beta$ was for binary covariates. Hence, the results of setting \textbf{1.} are very similar to the results described above. All tests increase in power for large sample sizes in both groups.  Again, the logrank test outperforms the  CGE-based tests in almost all settings. For unbalanced designs with lower $n_1$ and smaller covariates in group 1 (i.e. $\gamma < 0$), the same problems described for binary covariates with $\beta<0$ occur and the CGE tests perform worse than the logrank test. Again, the problems vanish, if both groups have a smaller sample size. These power properties are illustrated in Figure \ref{plot:power.unbalanced.normal} for the case of $\tau_{theor.}=0.25$ and $r=0.5$.

Setting \textbf{3.} has poisson distributed covariates with parameter $\lambda=1+\gamma$ and $\gamma$ being varied on $[-0.9,1.5]$ to consider various alternative hypothesis. Thus, mean and variance of the survival time generating covariates differ across groups. Still, all tests are able to maintain the type I error well and the power rises steadily with  $|\gamma|$ and achieves values close to one for large  $|\gamma|$ for all tests. For  sample sizes of $n_1=n_2=150$, this is illustrated in Figure \ref{plot:power.poisson}. In all settings, the logrank test's power is higher than that of  CGE-based tests, with power curves being clearly separated for all $\gamma$. For large sample sizes, the  CGE-based tests have an almost identical performance for all considered $\tau_{assum.}$.

Lastly, setting \textbf{2.} has normal covariates and alternative hypotheses of  standard deviation  1 in group 1 and varying deviation $\gamma$ in group 2. The mean of the covariates is the same for both groups.  CGE-based tests generally had higher power, since the logrank test was only  able to achieve acceptable power for large $n$ and low censoring rates. For $n_1=n_2=150$, dependency $\tau_{theor.}=0.25$ and a censoring proportion of 0.257 ($r=0.1$), the logrank test has a power estimate of 0.912 (SE of 0.912) at $\gamma =10$. At the same time, all four  CGE tests have a perfect power estimate of 1.000 (SE of 0.000). For a higher mean censoring proportion of 0.501, the  CGE-based tests lose power and e.g. have power estimates between 0.789 (SE of 0.013) and 0.892 (SE of 0.010) at $\tau_{theor.}=0.25$ and $\lambda=10$. The logrank test however only has a power of  0.382 (SE of 0.015) here.

\begin{figure}[h]
\centering
\includegraphics[scale=0.58]{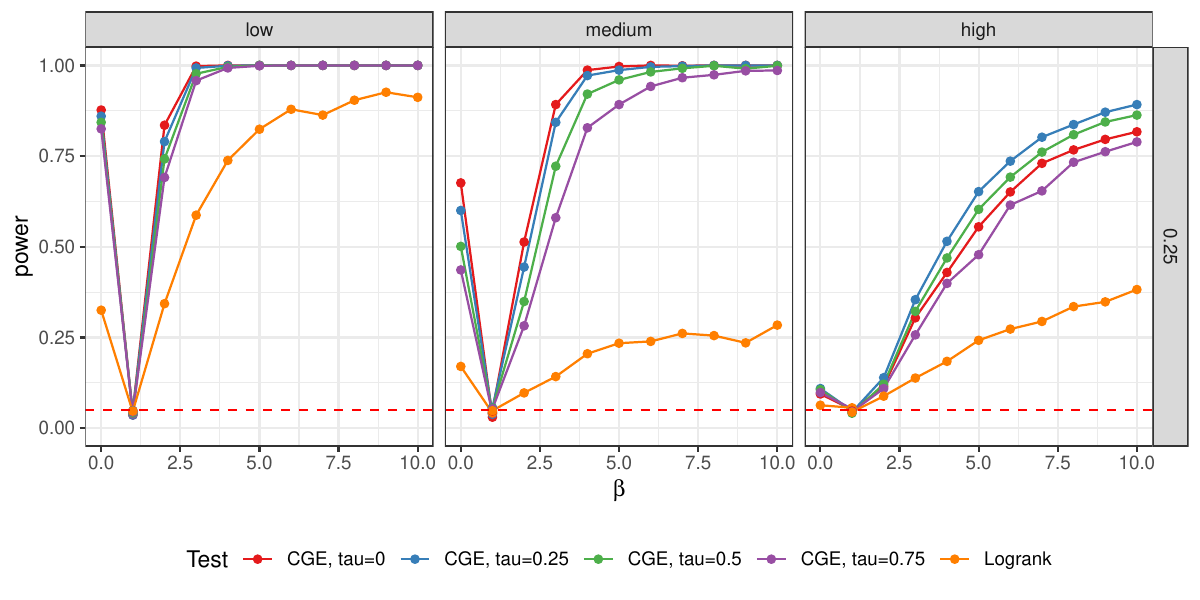}
\caption{Power estimates for normal covariates with varying standard deviation between groups, $n_1=n_2=150$ and $\tau_{theor.}=0.25$. Censoring percentages vary across columns from 0.258 and $r=0.1$ on the left to 0.501 and $r=0.5$ on the right. \label{plot:power.normal.var}}
\end{figure}

Setting \textbf{2.} is the only setting where  CGE-based tests with higher $\tau_{assum.}$ repeatedly show greater power than  the CGE test  with $\tau_{theor.}=0$, as can be seen in Figure \ref{plot:power.normal.var}. In the case of $n_1=n_2=50$, $\tau_{theor.}=0.5$ and $r=0.5$
the  CGE-based test with $\tau_{theor.}=0.75$ even has the highest power estimates, which we don't see for any other covariate distributions. See Figure \ref{plot:power.normal.var.tau} for details.   Looking at exemplary datasets from this settings, such as the one displayed in Figure \ref{plot:cge.ex.curve.var.normal}, we see that the longest observed times in both groups are similar for large $\gamma$. However, in group 2, survivals and censorings are completely seperated, meaning that all observed survival times are lower than any observed censoring time. Consequently, the curve of the CGE is at a relatively high level, since the censoring times do not affect the estimator's path at all.  The CGE of group 1 on the other hand, especially with large $\tau_{assum.}$,  strongly weights censorings in group 1 and falls to a much lower value. Thus, there is a good separation between the estimators of the groups, resulting in a small $p$-value. It should be noted that the  dataset from Figure \ref{plot:cge.ex.curve.var.normal} was an extreme example regarding the separation of survival and censoring times in group 2, which was not the case for all rounds of simulation.

\section{Simulation Study: Trees}\label{study2}
\subsection{Simulation Design}\label{sec:tree.study.design}

In a second simulation study, we compare the performance of the trees introduced in Section \ref{sec:survival.trees} to those of the logrank tree regarding prediction ability and ability to select relevant covariates. The study design is motivated by Emura et al. (2012)  \cite{emuraSurvivalPredictionBased2012} who generate datasets of survival times an according informative and non-informative covariates to study methods for compound covariate prediction. 

For each round of simulation, we simulate a training and a testing dataset with $n=100$ subjects each as follows: For subject $i$, the covariate vector 
\begin{equation*}
    \boldsymbol{z_i} = (   \underbrace{z_{i1}, \ldots, z_{iq}}_{\times q}, 
                            \underbrace{z_{i(q+1)}, \ldots,   z_{i(2q)}}_{\times q},
                            \underbrace{z_{i(2q+1)}, \ldots,   z_{ip}}_{\times p-2q}
                            ), \quad p,q \in \mathbb{N}, p>2q,
\end{equation*}

is drawn with each entry following a continuous uniform distribution with mean 0 and standard deviation 1.The covariates are separated into three blocks (1 to $q$, $q+1$ to $2q$ and $2q+1$ to $p$). Covariates of the first two blocks have a pair-wise correlation of $\rho$ within their block; the remaining genes are uncorrelated. We use the \texttt{X.pathway} function from the \texttt{R} - \texttt{compound.Cox} package to generate datasets of this format \cite{emuraCompoundCox2023}.  To mimic a clinical dataset consisting of categorical patient data and laboratory biomarker assessments, half of the genes in each block are transformed to a binary scale by calculating $\text{median}_i(z_{ij})$ for each coavriate $j \in \{1, \ldots, p\}$ and setting the new, binary covariate to $\tilde{z}_{ij} = \mathds{1}\{z_{ij} \geq   \text{median}_i(z_{ij}) \}$. Subsequently, all covariates are rounded to one digit after the decimal point, which reduces the size of possible cutoff-values within  survival trees and thus greatly speeds up the simulation study. Survival and censoring time are generated analogously to Section \ref{sec:simuation.design.test}, using a multidimensional parameter

\begin{equation*}
    \boldsymbol{\beta}   = (   \underbrace{\beta, \ldots, \beta}_{\times q}, 
                            \underbrace{-\beta, \ldots,  -\beta}_{\times q},
                            \underbrace{0, \ldots ,  0}_{\times p-2q}
                            )
\end{equation*}
to generate $T = -\log(U)\exp(-\boldsymbol{\beta}^\top\boldsymbol{z_i})$ for uniform $U$ and one-dimensional auxiliary parameter $\beta$. Note that only the first $2q$ covaraites actually affect the survival times. Again, the Clayton copula with parameter $\tau_{theor.}$ is used to generate dependency between survival and censoring times. We compare the performance of  CGE-based trees for $\tau_{assum.} \in \{  0.0001, 0.25,   0.5, 0.75 \}$  and the logrank tree using the implementation of the \texttt{uni.survival.tree} package \cite{emuraUniSurvivalTree2021}. We consider  larger sample sizes of  $n \in \{100, 300\}$, in order to have more subject observations than the number of covariates, which was chosen as  $p=50$. As this is only a small simulation study, all other parameter values are not varied, but approximately set in the middle of their respective feasible ranges, which leads to  $\beta = 0.5$, $\rho = 0.5$,   $p$-value threshold $\tilde{p} = 0.01$, and $\tau_{theor.} = 0.25$. Two censoring scenarios are considered leading to mean censoring proportion of 0.135 and 0.464, respectively. $n_{sim} = 100$ simulation rounds were performed and the number of permutations for the permutation tests was set to $n_{perm}=1000$, which allowes us to conduct the study in a manageable amount of time. The trees are evaluated regarding three criteria:

\textbf{1.} Selection ability: We evaluate, how many of the tree's splits are based on the informative covariates 1 to $2q$. We provide the precision, which is the proportion of non-terminal nodes that split depending on one of the informative covariates. It can take values between 0 and 1, with 1 indicating perfect precision  \cite{emuraSurvivalTreeBased2023}.

\textbf{2.} Prognostic ability:  We use Harrell's $C$-index to measure, if a tree's terminal nodes are able to order the test dataset's survival times correctly into groups from low to high survival. The terminal nodes are numerated from right to left with low numbers indicating low risk. For any two patients, the index describes, if the patient with the lower terminal node number survived longer than the patient with the higher  node number \cite{harrellEvaluatingYieldMedical1982}. The index is computed by a pairwise comparison over all test subjects, where one certainly had an event before the other one. If $(tn\#)_i$ denotes the terminal node number of subject $i$, with a low number indicating long survival, the numbers of concordant (CC), discordant (DC) and tied (TR) pairs of all comparable subjects $i,j = 1, \ldots, n$ are given by \cite{bertsimasOptimalSurvivalTrees2022}:
\begin{align*}
    CC &= \sum\limits_{i,j} \mathds{1}\{x_i>x_j\} \mathds{1}\{(tn\#)_i<(tn\#)_j\}, \,
    DC = \sum\limits_{i,j} \mathds{1}\{x_i>x_j\} \mathds{1}\{(tn\#)_i>(tn\#)_j\}, \, \\
    TR &= \sum\limits_{i,j} \mathds{1}\{x_i>x_j\} \mathds{1}\{(tn\#)_i=(tn\#)_j\}.
\end{align*}
The resulting index
\begin{equation*}
    H_C = \frac{CC+0.5TR}{CC+DC+TR}
\end{equation*}
can take values in $[0,1]$, with high values indicating good prognostic ability  \cite{bertsimasOptimalSurvivalTrees2022}. The \texttt{survival} package was used to calculate Harrell's $C$ \cite{survival-package}.

\textbf{3.} Prediction ability: The Brier Score is used to evaluate, how accurately a tree predicts the survival of a subject from the testing dataset at a given timepoint. We predict the survival of a subject in terminal node $(tn\#)$ at time $t$   as the value of the  CGE calculated on subjects in node  $(tn\#)$ from the training dataset at time $t$. We calculate the  CGE using the same $\tau_{assum.}$ used to construct the respective tree, and use $\tau_{assum.}\approx 0$, i.e. the Kaplan-Meier estimator for the logrank tree predictions.  Graf et al. (1999) \cite{grafAssessmentComparisonPrognostic1999}  provide a version of the Brier score that incorporates censoring by weighting deviations of predicted survival from true survival with an estimator of the censoring distribution. This score can then be integrated over the study period to yield the Integrated Brier Score, which is

\begin{equation}\label{eq:ibs}
IB = \frac{1}{\max(\boldsymbol{x})}
    \frac{1}{n} \sum\limits_{i=1}^n 
        \left\{
        \int\limits_0^{x_i}
        \frac{\left(1- \hat{S}(t|\boldsymbol{z_i})\right)^2}{\hat{C}(t)} dt
        +
        \delta_i \int\limits_{x_i}^{\max(\boldsymbol{x})}
            \frac{\hat{S}(t|\boldsymbol{z_i})^2}{\hat{C}(x_i)} dt         
        \right\},
\end{equation}
with $\hat{S}$ and $\hat{C}$ describing estimators of survival or censoring function, respectively \cite{bertsimasOptimalSurvivalTrees2022}. Common choices are Kaplan-Meier estimators. We  additionally provide the Integrated Brier Score using  CGEs with  $\tau_{assum.}=\tau_{theor.}$, which in the case of this simulation study is known to us. The integral from Equation \eqref{eq:ibs} is estimated over a grid of timepoints $\tilde{t}_1, \ldots, \tilde{t}_m$ as

\begin{equation*}
    \widehat{IB} = \frac{1}{\max(\boldsymbol{x})}
        \sum\limits_{j=1}^{m-1} \left\{ (\tilde{t}_{j+1} - \tilde{t}_j)
                \frac{1}{n} \sum\limits_{i=1}^n 
                    \left\{
                         \frac{\left(1- \hat{S}(\tilde{t}_j|\boldsymbol{z_i})\right)^2}{\hat{C}(\tilde{t}_j)}\mathds{1}\{x_i\geq\tilde{t_j}\}
                    +
                        \frac{\hat{S}(\tilde{t}_j|\boldsymbol{z_i})^2}{\hat{C}(x_i)} \mathds{1}\{x_i<\tilde{t_j}\}
                    \right\} \right\}.
\end{equation*}
It is sufficient to choose  $\tilde{t}$ as the timepoints observed in the testing or training dataset, since the step-function like estimators will not fluctuate in between these points. \footnote{\texttt{R} code for the estimator $ \widehat{IB}$ is taken from the package \texttt{SurvMetrics} \cite{zhouSurvMetricsPredictiveEvaluation2022}, where Kaplan-Meier estimators were replaced with  CGEs. Instead of considering step-functions, \texttt{SurvMetrics} interpolates the Kaplan-Meier estimate for the censoring function inbetween censoring times (see code of fucntion \texttt{Gt()} for details). In this paper,  all estimators are step functions. Thus, our estimate of $IB$ for $\tau_{theor.} \approx 0$ is only identical to \texttt{SurvMetrics}'s estimate, if the $\tilde{t}$ are exactly the timepoints were a censoring was observed. However, the deviations between the methods are negligibly small.}

\subsection{Simulation Results}\label{sec:study.tree.res}

Table \ref{tab:tree:study:result} displays the study results. The performance of the four  CGE-based trees is very similar in all settings regarding the number of terminal nodes as well as all accuracy measures introduced in Section  \ref{sec:tree.study.design}. No noticeable  trends related to $\tau_{assum.}$  can be seen.

The logrank trees have more terminal nodes than the  CGE trees. For instance, the  CGE tree with  $\tau_{assum.}=0.25$ has 15.7 terminal nodes on average for $n=100$ and low censoring of $0.135$\%, while the logrank tree has 24.450, which is the 1.557-fold. As $n$ rises from 100 to 300, the number of terminal nodes increases with a factor of about 3 for the  CGE trees (e.g. factor 2.926 for $\tau_{assum.}=0.5$ and low censoring) and a  factor around 2.7 for the logrank tree (2.692 for low censoring). These large numbers for $n=300$, especially seen for the logrank trees,  might diminish the easy and practical interpretability of survival trees.  Adjusting the threshold $p$-value $\tilde{p}$ or adding a minimum nodesize threshold to allow an additional split to the tree algorithm would solve this problem. Lastly, the trees are smaller for higher censoring, which aligns with the results from Section \ref{sec:test.study.result}, where we saw that both logrank and  CGE-based tests lose power with higher censoring and therefore find fewer significant splits. 

The precision of the  CGE trees is over 0.666 in all settings. Hence, the majority of covariates is selected correctly. The logrank tree's precision is lower in all settings.  For all trees, the precision decreases with higher $n$ (e.g. from 0.762 for $n=100$ to 0.681 for $n=300$ for $\tau_{assum.}=0.5$), which might be partially caused by a higher number of cutoff values and thus the higher number of possible splits that comes with larger $n$.  In all settings except for $n=100$ and low censoring, the  CGE test with $\tau_{assum.}=0$ has the largest precision ranging between 0.682 and 0.799. However, the differences are small (e.g. 0.710 for $\tau_{assum.}=0$  vs. 0.681  for $\tau_{assum.}=0.5$ for $n=300$ and high censoring). 

Both the traditional Brier Score and the  CGE adjusted score have higher values for higher censoring, indicating less prediction ability. For instance, the score increases from 0.120 for low censoring to 0.288 for high censoring for the  CGE test with $\tau_{assum.}=0.25$  and $n=100$. Between $n=100$ and $n=300$ only minor changes in the Kaplan-Meier estimate based Integrated Brier Score can be seen. In contrast, the  CGE score surprisingly increases, e.g. from 0.354 to   0.422 for low censoring and the test with $\tau_{assum.}=0.75$. This observation calls for further investigation into the properties of the  CGE-based Brier Score.  In almost settings, the logrank tests shows the highest values for both versions of the Integrated Brier Score.

\begin{table}[t]
\centering
\caption{Average performance measures from 100 rounds of simulation. $H_C$ denotes Harrell's $C$-index, $\widehat{IB}$ KM is the  Integrated Brier Score and $\widehat{IB}$ CGE the Integrated Brier Score with the  CGE with $\tau = 0.25$. Mean censoring proportions of 0.135 (low) and 0.464 (high)} \label{tab:tree:study:result}
\begin{tabular}{rrcrrrrr}
  \hline
 &&& \multicolumn{4}{c}{Copula-graphic estimator} & \\
 &&& $\tau_{assum.}=0$ & $\tau_{assum.}=0.25$ & $\tau_{assum.}=0.5$ &  $\tau_{assum.}=0.75$ & Logrank \\ 
  \hline
 $n=100$ &low cens. & $\#$ term. nodes & 15.800 & \textbf{15.700} & 15.760 & 15.920 & 24.450 \\ 
 & &Precision\% & 0.765 & 0.762 & \textbf{0.776} & 0.752 & 0.619 \\ 
 & &$H_C$ & 0.715 & 0.712 & 0.711 & 0.715 & \textbf{0.718} \\ 
 & &$\widehat{IB}$ KM & 0.127 & 0.120 & 0.118 & \textbf{0.115} & 0.241 \\ 
&&$\widehat{IB}$ CGE  & 0.358 & \textbf{0.336} & 0.356 & 0.354 & 0.420 \\  \hline

& high cens.   & $\#$ term. nodes& \textbf{13.330} & 13.480 & 13.550 & 13.750 & 17.660 \\ 
 &  &Precision\%  & \textbf{0.799} & 0.789 & 0.762 & 0.778 & 0.650 \\ 
 & &$H_C$ & 0.730 & \textbf{0.736} & 0.732 & 0.733 & 0.728 \\ 
 & &$\widehat{IB}$ KM  & \textbf{0.286} & 0.288 & 0.296 & 0.286 & 0.340 \\ 
 &  &$\widehat{IB}$ CGE & \textbf{0.459} & 0.492 & 0.531 & 0.528 & 0.526 \\ \hline \hline

 $n=300$&low cens.   &$\#$ term. nodes & 46.520 & 46.410 & \textbf{46.120} & 46.480 & 65.810 \\ 
  & &Precision\%  & \textbf{0.682} & 0.681 & 0.669 & 0.666 & 0.529 \\ 
  &&$H_C$ & 0.712 & 0.715 & 0.712 & 0.715 & \textbf{0.732} \\ 
  &&$\widehat{IB}$ KM & 0.109 & 0.108 & \textbf{0.101} & 0.107 & 0.214 \\ 
& &$\widehat{IB}$ CGE & 0.396 & 0.420 & \textbf{0.391} & 0.422 & 0.521 \\ \hline

 & high cens. &  $\#$ term. nodes& 39.280 & 38.760 & \textbf{38.580} & 38.810 & 46.300 \\ 
 & &Precision\% & \textbf{0.710} & 0.680 & 0.681 & 0.674 & 0.557 \\ 
 &&$H_C$  & 0.740 & 0.738 & 0.742 & 0.744 & \textbf{0.753} \\ 
 & &$\widehat{IB}$ KM  & 0.289 & 0.291 & \textbf{0.283} & 0.292 & 0.355 \\ 
 &  &$\widehat{IB}$ CGE & \textbf{0.511} & 0.555 & 0.566 & 0.576 & 0.601 \\ 
   \hline \hline
\end{tabular}
\end{table}

For Harrell's $C$-index, the row-wise values of all trees are comparable. In particular, the  logrank trees have the highest $C$-index in three out of four settings (for instance a value of 0.732 for $n=300$ and low censoring). All $C$-indices are well over $0.500$, which would indicate  trees that order subjects  no better than random assignment. The $C$-indices are slightly higher in the high censoring scenario, but do not change much with $n$.

\section{Illustrative Data Analysis}\label{example}
\label{sec:data_example}
In this section, we will apply the proposed tree algorithm to real-world data. We will use the Primary Biliary Cholangitis (PBC) dataset provided by the \texttt{survival} package \cite{survival-package}.  PBC is a progressive liver disease causing inflammations in the liver that lead to cirrhosis, destruct the bile ducts and eventually result in   death. The dataset is from a randomized Mayo Clinic trial conducted between 1974 and 1984 testing  D-penicillmain, a possible treatment for PBC   \citep[Chapter 3]{therneauCoxModel2000}. 

The full dataset  contains data from 418 subjects. However, 106 did not participate in the randomized trial and consequently have many missing values in variables potentially very relevant to survival, in particular the variable \textit{Treatment}. Thus, these 106 subjects are removed prior to our data analysis, leaving us with a sample size of 312.

In addition to each patient's event or censoring time, 17 covariates are provided. These include seven binary or categorical variables  such as \textit{Presence of Ascites} and \textit{Presence of Hepatomegaly or Enlarged Liver}. Furthermore, ten continuous variables, mostly biomarkers, such as \textit{Serum Bilirunbin} or \textit{Serum Cholesterol}  are provided. Before analysis, we rounded the variable \textit{Age} to full years to reduce the value of feasible cutoff values and save computation time.

1.080\% of covariate observations are missing, especially of the variables \textit{Serum Cholesterol} (8.974\% missing) and \textit{Triglycerides} (9.615\% missing). So far, the proposed tree cannot handle missing values. Consequently, these two variables were removed from the dataset prior to our analysis. After this, only six subjects with missing values in one or more varialbes were left. These subjects were removed as well, leaving 306 subjects for the following analysis.

Of the remaining patients, 123 died during the study, 164 were censored due to lost follow-up and 19 had a liver transplantation. For this illustrative data example, these patients are considered censored as well, since the transplantation prevented a death through PBC. 

It is reasonable to assume, that a patient's health status, and with it their underlying survival time distribution, affected the decision, if a liver transplantation was necessary \cite{staplinDependentCensoringPiecewise2015}. Thus, we suspect that the dataset was generated by positively dependent survival and censoring times, making it a suitable dataset to test our method.

Before constructing a survival tree, we evaluate the performance of the tree construction algorithms by cross-validation. We evaluate the performance of the CGE trees using the Clayton copula with Kendall's  $\tau_{assum} \in \{0.0001, 0.125, 0.250, 0.375, 0.500, 0.625, 0.750, 0.875\}$ and the logrank tree. We use a $p$-value threshold of $\tilde{p}$=0.01, the number of permutations $n_{perm.}=5000$ and apply the metrics introduced in Section \ref{sec:tree.study.design} to evaluate the tree algorithms. The performance measures are computed using 5-fold cross-validation \citep[Chapter 7]{hastieModelAssessmentSelection2009}.

 The results can be seen in Table \ref{tab:result.data.example}. The logrank tree has  23.8 terminal nodes on average, which is slightly more than the CGE trees, which have between 14.6 terminal nodes for $\tau_{assum.}=0.875$ and 21.0 terminal nodes for  $\tau_{assum.}=0.125$. The larger size of the logrank tree was also seen in the simulation study in Section \ref{sec:study.tree.res}. The size of the CGE trees tend to decrease with  $\tau_{assum.}$.  At the same time, the mean $C$-index increases almost monotonously from 0.718 at $\tau_{assum.}=0.000$ to 0.756 at $\tau_{assum.}=0.875$. The logrank tree had the highest $C$-index of 0.800.

\begin{table}[h]
\centering
\caption{Mean performance measures on PBC data over 5-fold cross validation. $H_C$ denotes Harrell's $C$-index and $\widehat{IB}$ KM is the Integrated Brier Score based on the Kaplan-Meier estimator. The mean censoring proportion is 0.598.}
\label{tab:result.data.example}

\begin{tabular}{rrrrrrrrrr}
  \hline
  \hline
Tree &Logrank & \multicolumn{8 }{c}{$\tau_{assum.}$ of CGE } \\
 &  & 0.000 & 0.125 & 0.250 & 0.375 & 0.500 & 0.625 & 0.750 & 0.875 \\ 
  $\#$ term. nodes & 23.8 & 18.6 & 21.0 & 18.0 & 17.0 & 17.4 & 16.2 & 18.0 & \textbf{14.6} \\ 
 $\widehat{IB}$  KM& 0.210 & \textbf{0.184} & 0.188 & 0.197 & 0.208 & 0.204 & 0.214 & 0.206 & 0.213 \\ 
  $H_C$ & \textbf{0.800} & 0.716 & 0.725 & 0.732 & 0.744 & 0.742 & 0.750 & 0.752 & 0.756 \\ 
   \hline
\end{tabular}
\end{table}

The Kaplan-Meier-based Integrated Brier Score rises with $\tau_{assum.}$. The CGE trees with  $\tau_{assum.}=0.0625$ and  $\tau_{assum.}=0.875$ have a higher score than the logrank tree. All other CGE trees have a lower score with values below 0.208. 

The analysis of performance measures does not indicate one CGE tree that is superior to the other ones. This result is confirmed by a visualization of the CGE-based tree's metrics in Figure \ref{fig:ex.measure} in the Appendix. In the following, we will choose the tree with $\tau_{assum.}=0.375$ for a more detailed analysis, since it provides a compromise of relatively small mean tree size, large $C$-index and an Integrated Brier Score smaller than the logrank tree's. We will compare this tree's properties to that of the logrank tree. Both trees are re-calculated on the full dataset.

\begin{figure}[h]
\centering
\includegraphics[scale=0.4]{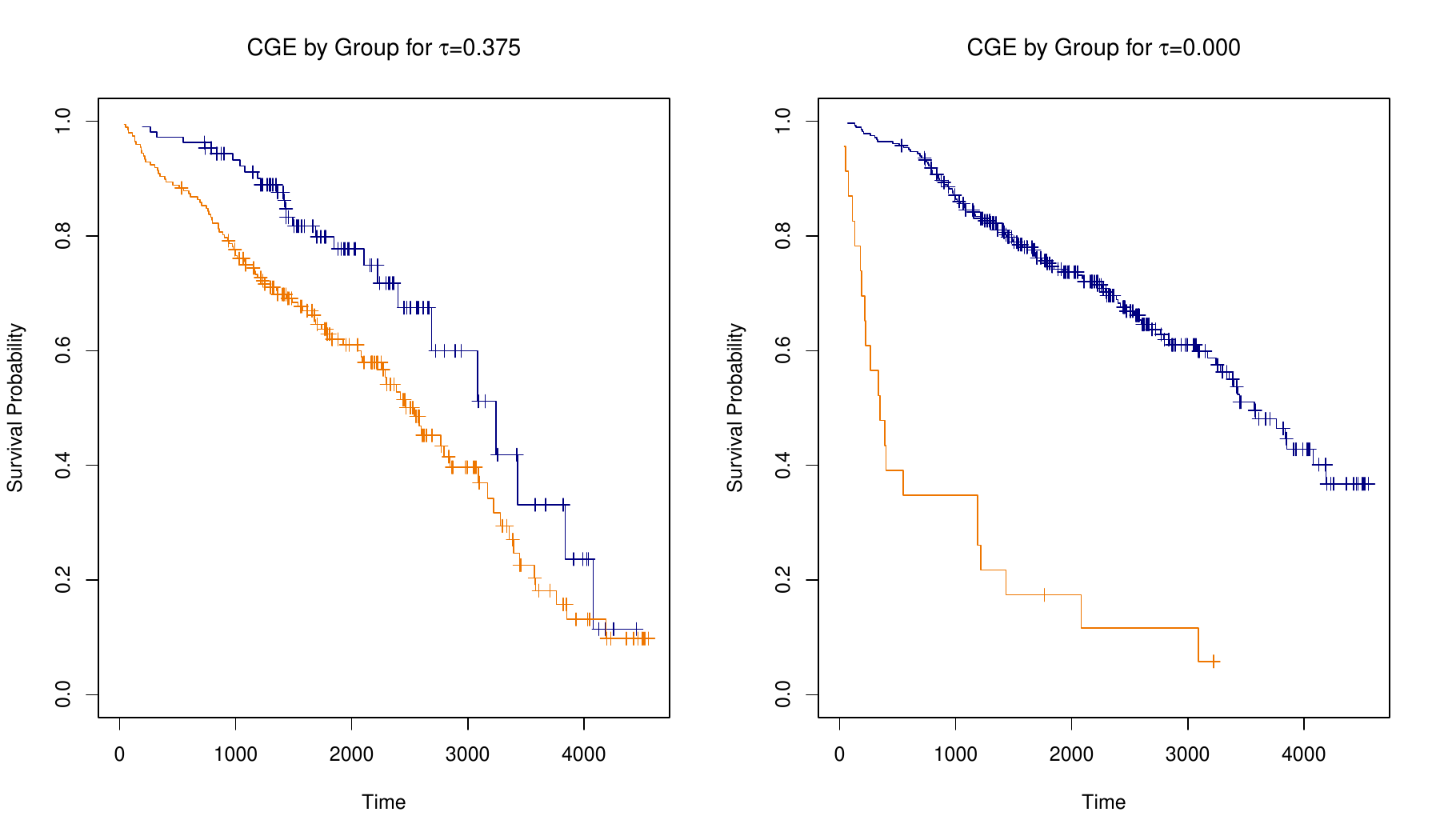}
\caption{Survival curve estimates after the first split  for the CGE tree with $\tau_{assum.}=0.375$  (left) and the logrank tree using the Kaplan-Meier estimator (right).   \label{fig:first_split1}}
\end{figure}

The resulting trees are again larger than the ones seen during cross validation due to increased $n$. This results in a CGE tree with 22 and a logrank tree with 30 terminal nodes. Both trees seem overfitted. Half of the CGE tree's terminal nodes have under  five subjects. The logrank tree has 18 terminal nodes with under five  and ten with only one subject. No clear separation of the survival curves of the terminal nodes was visible and patterns in survival across terminal nodes could not be seen clearly. See Figure  \ref{fig:terminal_node}  in the Appendix for details. Both trees use multiple varialbes repeatedly for splitting in subsequent splits. We thus re-calculate both trees using a smaller $p$-value threshold of $\tilde{p}=0.001$. The new trees were considerably smaller with 12 terminal nodes for the CGE tree and 19 terminal nodes for the logrank tree. We will describe these trees for the remaining part of this section:

The separate group survival curve estimates at the first split of both trees can be seen in Figure \ref{fig:first_split1}. The CGE tree splits the data into $\{\textit{Age} \leq 45\}$ and $\{\textit{Age} > 45\}$ , with the first group showing longer survival. Figure \ref{fig:first_split1} indeed shows a clear separation of the CGEs of the two group's survival, with the older group having the lower estimates.   The first split of the logrank tree is by $\{\textit{Presence of Ascites} > 0\}$ and $\{\textit{Presence of Ascites} \leq 0\}$. The separated groups by the logrank test are uneven, with only 23 of 306 subjects being in the second group. Moreover, the censored subjects are distributed among groups very unevenly with only two of 183 censored subjects being in the group with lower expected survival.

\begin{figure}
\centering
\includegraphics[scale=0.342]{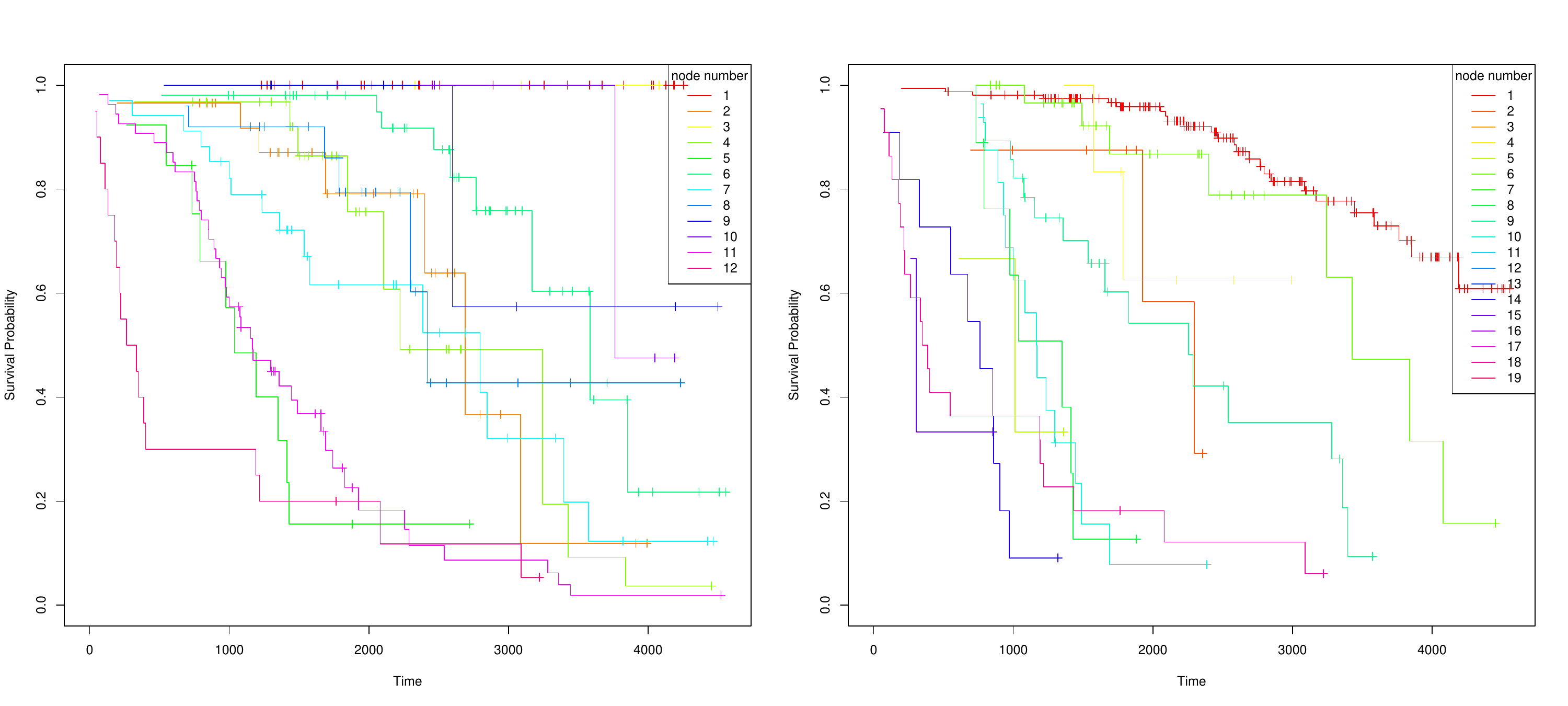}
\caption{Survival curve estimates of terminal nodes, with node 1 indicating highest survival probability. CGE tree with $\tau_{assum.}=0.375$  (left) and the logrank tree using the Kaplan-Meier estimator (right). The trees were calculated with $\tilde{p}=0.001$. High node numbers indicate shorter survival. \label{fig:terminal_node2}}
\end{figure}

We continue our analysis by looking at the  survival curve estimators of the PBC data by terminal node in Figure \ref{fig:terminal_node2}. Still, no clear separation of estimates is visible, but some tendency for nodes with lower numbers having longer survival times is visible. The graphic for the CGE tree shows several groups that only consist of censored individuals with overlapping CGEs constantly at 1. We can also see that the terminal node sizes for the CGE tree are distributed relatively evenly by taking values between 6 and 54. The logrank tree's final node sizes differ more from each other. One terminal node contains 157, more than half, of the study subjects. Eight terminal nodes only contain one subject, and their survival estimator cannot be displayed in Figure \ref{fig:terminal_node2}. The more even distribution of subjects over terminal nodes speaks in favor of the CGE tree here.

\begin{figure}
\centering
\includegraphics[scale=0.8]{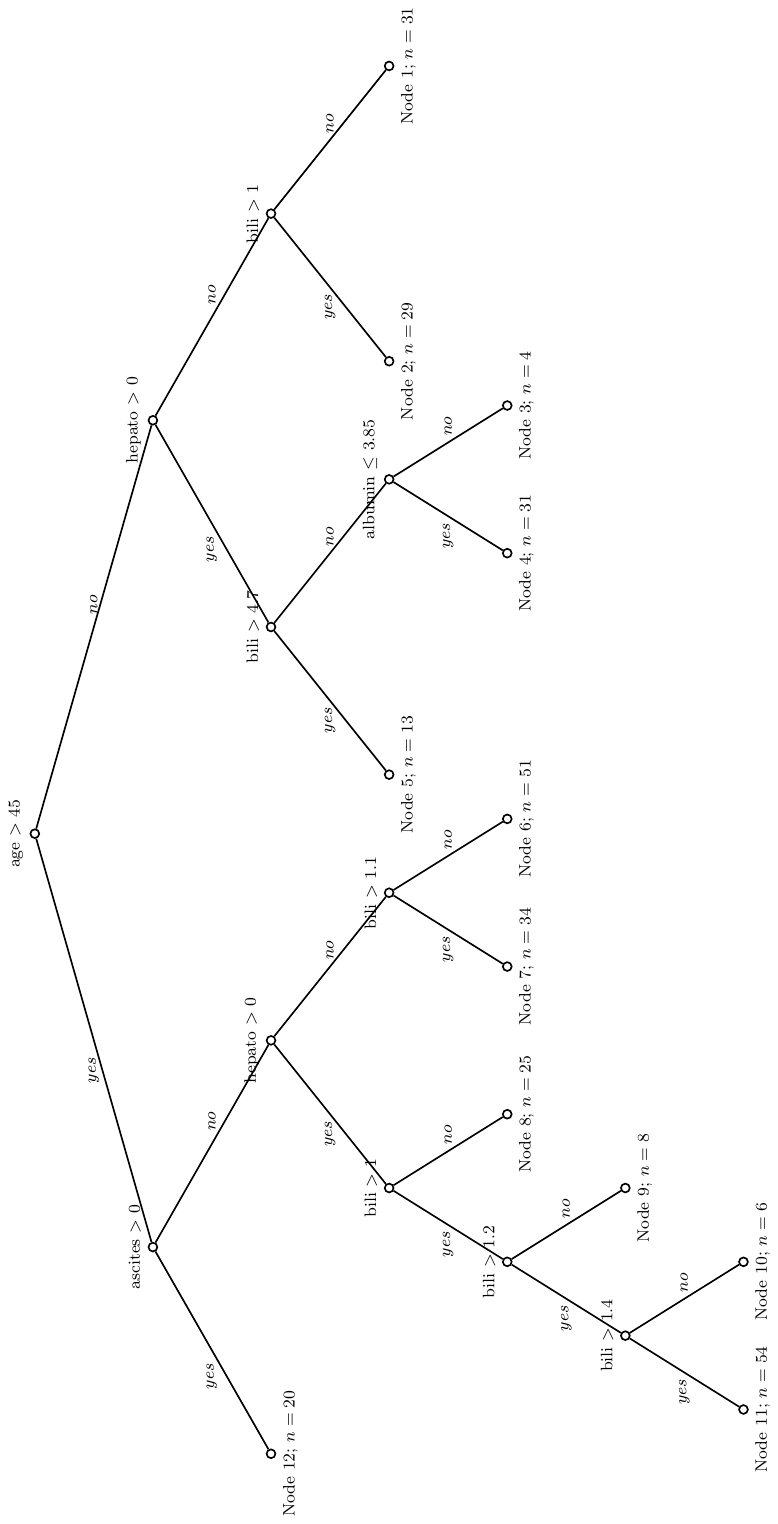}
\caption{CGE survival tree with $\tau_{assum.}=0.375$ and $\tilde{p}=0.001$. High node numbers indicate shorter survival. \label{fig:tree.full}}
\end{figure}

To end our analysis, we display the full CGE classification tree for $\tau_{assum.}=0.375$  in Figure \ref{fig:tree.full}. Many of the tree's splits seam reasonable, such as categorizing older patients (\textit{Age} $>$45) or patients with \textit{Presence of Hepatomegaly or Enlarged Liver} (hepto) in higher risk groups. Nine of the twelve terminal nodes have a parent node that splits according to \textit{Serum Bilirunbin}. While this points out the importance of the biomarker for liver health, it also indicates that the tree might be overfitted. In further analysis, we could for instance find a systematic way to summarize node 8 to 11, which result from repeatedly splitting the data according to \textit{Serum Bilirunbin}.

\newpage

\section{Discussion}

In this paper, we introduced a permutation test for the null hypothesis of equal survival distributions in two groups under the assumption of equal censoring distributions. The introduced test is based on the copula-graphic estimator, such that form and strength of a dependence between survival and censoring times can be incorporated. Assuming a Clayton copula and various dependence parameters corresponding to values of Kendall's $\tau$ between 0 and 0.75, we assessed the test's type I error and power in a simulation study. We found that the tests maintains a desired type I error of $\alpha=0.05$ relatively well for all considered sample sizes and censoring scenarios. Furthermore, it was robust to a misspecification of the copula model when the data was generated from a Frank copula. 
We were overall satisfied with the tests' power, which was generally robust except in scenarios with highly unbalanced group sizes and significant censoring. Here, the logrank test demonstrated higher power than our proposed test, even under conditions of dependent censoring. The most notable power advantage of our proposed test compared to the logrank test was observed in cases with heteroscedasticity.

The permutation test was then implemented as a splitting criterion in a survival tree algorithm, that was tested on simulated data as well as on real-world data from the Mayo Clinic Primary Biliary Cholangitis clinical trial. While the trees' survival predictions demonstrated satisfactory concordance in all settings, with Harrell's $C$-indices exceeding than 0.7, the Integrated Brier Score was rather high with values of over 0.28, especially for data with a high censoring rate above 46\%.  

One issue that was especially apparent in the data example, was a high number of terminal nodes. This number was often larger than the number of covariates and took away some of the interpretability of our trees that we had hoped for. Therefore, some adjustments to our tree algorithm should be made in the future. We could implement a minimum nodesize threshold that prevents splits when only few subjects are left in a node. We also saw a problem of copula-graphic estimator trees finding splits with large absolute differences between the group's estimators by choosing one group consisting of almost exclusively censored subjects. This resulted into one group's CGE having values close to 1 over the whole study period and led to large test statistic values. To prevent splits like the one seen in Figure \ref{fig:second_split}, we could add a maximum censoring threshold for each descendant group. Furthermore, we could add an amalgamation step to our algorithm that reduces the number of terminal nodes by recursively grouping similar terminal nodes together  \cite{ciampiTreestructuredPredictionCensored1995, bou-hamadReviewSurvivalTrees2011}. 
However, 
the logrank tree had even more terminal nodes than the copula-graphic estimator based trees across all study settings. One reason for this is given by our $p$-value threshold of $0.01$ that was set for all trees before the start of the simulation study. A detailed study on its calibration for both types of trees would be necessary to properly evaluate this issue.

During our study, we encountered very long computation times for survival trees, caused by the nesting of the iterative permutation test into the iterative tree algorithm. To solve this problem, we could try to adapt a computationally efficient, matrix-based algorithm \cite{emuraSurvivalTreeBased2023} for a survival tree splitting based on score tests.  While a direct transfer of this algorithm to our proposed test does not seem possible, since we use a permutation test, some modification of the proposed algorithm might work for our setting.

One of the major limitations of the proposed method is the assumption of equal censoring distributions of both groups. While we need this assumption and the resulting exchangeability of survival times under the null hypothesis for now to mathematically derive an exact test, it is not a realistic assumption and limits the use of our work on real-world data. Other authors  \cite{brendelWeightedLogrankPermutation2014,ditzhausMorePowerfulLogrank2020} suggest an alternative permutation approach that does not require exchangeability under the null hypothesis by applying a studentization. They studentize their test statistic by adding a (co)variance estimator to the test statistic, that is calculated on the same data permutations as the rest of the statistic. This studentized permutation strategy was subsequently applied to inference in factorial designs  \cite{ditzhausCASANOVAPermutationInference2023}.  A similar studentization approach could also be derived for our statistic.
Moreover, we could also think about other tests for splitting, e.g. based on adoptions of ideas from omnibus tests \cite{brendelWeightedLogrankPermutation2014,ditzhausMorePowerfulLogrank2020,dormuthComparativeStudyAlternatives2023} or based on interpretable estimands like the restricted mean survival time \cite{ditzhausStudentizedPermutationMethod2023}. Lastly, we decided to only consider a single survival tree to benefit of its straightforward interpretability and low computation time. However,  tree ensembles such as survival forests \cite{ishwaran} are shown to have much higher predictive ability. Furthermore, during 5-fold cross validation we noticed differences in the five resulting trees, their number of terminal nodes and chosen covariates. This instability could also be reduced using an ensemble method. 
Although a survival forest may increase computational time and reduce interpretability, its potential predictive benefits could justify the trade-off. Beyond the classical random survival forest \cite{ishwaran}, many other implementations exist, see for example Section 3 of Bou-Hamad et al. (2011)  \cite{bou-hamadReviewSurvivalTrees2011} for an overview.


\section*{Acknowledgments}
The authors gratefully acknowledge the computing time provided on the Linux HPC cluster at Technical University Dortmund (LiDO3), partially funded in the course of the Large-Scale Equipment Initiative by the Deutsche Forschungsgemeinschaft (DFG, German Research Foundation) as
project 271512359. Moreover, Markus Pauly was funded by the Deutsche Forschungsgemeinschaft (DFG, German Research Foundation) project 352692197. 

We occasionally used the ChatGPT 4.0 model from OpenAI for minor language edits, aiming to enhance readability. After using this tool, the authors reviewed and edited the content as needed and take full responsibility for the content.



\section*{Conflict of interest}

The authors declare no potential conflict of interests.

\newpage

\bibliography{literature.bib}

\begin{thebibliography}{55}
\providecommand{\natexlab}[1]{#1}
\providecommand{\url}[1]{\texttt{#1}}
\expandafter\ifx\csname urlstyle\endcsname\relax
  \providecommand{\doi}[1]{doi: #1}\else
  \providecommand{\doi}{doi: \begingroup \urlstyle{rm}\Url}\fi

\bibitem[Leung et~al.(1997)Leung, Elashoff, and Afifi]{leungCensoringIssuesSurvival1997}
Kwan-Moon Leung, Robert~M. Elashoff, and Abdelmonem~A. Afifi.
\newblock Censoring issues in survival analysis.
\newblock \emph{Annual Review of Public Health}, 18\penalty0 (1):\penalty0 83--104, 1997.
\newblock ISSN 0163-7525.
\newblock \doi{10.1146/annurev.publhealth.18.1.83}.

\bibitem[Kaplan and Meier(1958)]{kaplanNonparametricEstimationIncomplete1958}
E.~L. Kaplan and Paul Meier.
\newblock Nonparametric {{Estimation}} from {{Incomplete Observations}}.
\newblock \emph{Journal of the American Statistical Association}, 53\penalty0 (282):\penalty0 457--481, 1958.
\newblock ISSN 0162-1459.
\newblock \doi{10.2307/2281868}.

\bibitem[Huang and Zhang(2008)]{huangRegressionSurvivalAnalysis2008}
Xuelin Huang and Nan Zhang.
\newblock Regression survival analysis with an assumed copula for dependent censoring: A sensitivity analysis approach.
\newblock \emph{Biometrics}, 64\penalty0 (4):\penalty0 1090--1099, 2008.
\newblock ISSN 1541-0420.
\newblock \doi{10.1111/j.1541-0420.2008.00986.x}.

\bibitem[Emura and Chen(2016)]{emuraGeneSelectionSurvival2016}
Takeshi Emura and Yi-Hau Chen.
\newblock Gene selection for survival data under dependent censoring: {{A}} copula-based approach.
\newblock \emph{Statistical Methods in Medical Research}, 25\penalty0 (6):\penalty0 2840--2857, 2016.
\newblock ISSN 0962-2802.
\newblock \doi{10.1177/0962280214533378}.

\bibitem[Klein and Moeschberger(1987)]{kleinIndependentDependentCompeting1987}
John~P. Klein and M.L Moeschberger.
\newblock Independent or dependent competing risks: Does it make a difference.
\newblock \emph{Communications in Statistics - Simulation and Computation}, 16\penalty0 (2):\penalty0 507--533, 1987.
\newblock ISSN 0361-0918, 1532-4141.
\newblock \doi{10.1080/03610918708812602}.

\bibitem[Schneider et~al.(2023)Schneider, {dos Reis}, Gottselig, Fisch, Knauth, and Vigo]{schneiderClaytonCopulaSurvival2023}
Silvana Schneider, Rodrigo Citton~P. {dos Reis}, Maicon M.~F. Gottselig, Patr{\'i}cia Fisch, Daniela~Riva Knauth, and {\'A}lvaro Vigo.
\newblock Clayton copula for survival data with dependent censoring: {{An}} application to a tuberculosis treatment adherence data.
\newblock \emph{Statistics in Medicine}, 42\penalty0 (23):\penalty0 4057--4081, 2023.
\newblock ISSN 1097-0258.
\newblock \doi{10.1002/sim.9858}.

\bibitem[Staplin et~al.(2015)Staplin, Kimber, Collett, and Roderick]{staplinDependentCensoringPiecewise2015}
N.~D. Staplin, A.~C. Kimber, D.~Collett, and P.~J. Roderick.
\newblock Dependent censoring in piecewise exponential survival models.
\newblock \emph{Statistical Methods in Medical Research}, 24\penalty0 (3):\penalty0 325--341, 2015.
\newblock ISSN 1477-0334.
\newblock \doi{10.1177/0962280214544018}.

\bibitem[Templeton et~al.(2020)Templeton, Amir, and Tannock]{templetonInformativeCensoringNeglected2020}
Arnoud~J. Templeton, Eitan Amir, and Ian~F. Tannock.
\newblock Informative censoring \textemdash{} a neglected cause of bias in oncology trials.
\newblock \emph{Nature Reviews Clinical Oncology}, 17\penalty0 (6):\penalty0 327--328, 2020.
\newblock ISSN 1759-4782.
\newblock \doi{10.1038/s41571-020-0368-0}.

\bibitem[Nelsen(2006)]{nelsenIntroductionCopulas2006}
Roger~B. Nelsen.
\newblock \emph{An {Introduction} to {Copulas}}.
\newblock Springer, New York, NY, 2006.
\newblock ISBN 978-0-387-28659-4.
\newblock \doi{10.1007/0-387-28678-0}.
\newblock URL \url{http://link.springer.com/10.1007/0-387-28678-0}.

\bibitem[Zheng and Klein(1995)]{zhengEstimatesMarginalSurvival1995}
Ming Zheng and John~P. Klein.
\newblock Estimates of {{Marginal Survival}} for {{Dependent Competing Risks Based}} on an {{Assumed Copula}}.
\newblock \emph{Biometrika}, 82\penalty0 (1):\penalty0 127--138, 1995.
\newblock ISSN 0006-3444.
\newblock \doi{10.2307/2337633}.

\bibitem[Lo and Wilke(2010)]{loCopulaModelDependent2010}
Simon M.~S. Lo and Ralf~A. Wilke.
\newblock A {{Copula Model}} for {{Dependent Competing Risks}}.
\newblock \emph{Journal of the Royal Statistical Society. Series C (Applied Statistics)}, 59\penalty0 (2):\penalty0 359--376, 2010.
\newblock ISSN 0035-9254.

\bibitem[Yeh et~al.(2023)Yeh, Liao, and Emura]{yehSensitivityAnalysisSurvival2023}
Chih-Tung Yeh, Gen-Yih Liao, and Takeshi Emura.
\newblock Sensitivity {{Analysis}} for {{Survival Prognostic Prediction}} with {{Gene Selection}}: {{A Copula Method}} for {{Dependent Censoring}}.
\newblock \emph{Biomedicines}, 11\penalty0 (3):\penalty0 797, March 2023.
\newblock ISSN 2227-9059.
\newblock \doi{10.3390/biomedicines11030797}.

\bibitem[Braekers and Veraverbeke(2005)]{braekersACopula2005}
Roel Braekers and Noël Veraverbeke.
\newblock A copula-graphic estimator for the conditional survival function under dependent censoring.
\newblock \emph{The Canadian Journal of Statistics / La Revue Canadienne de Statistique}, 33\penalty0 (3):\penalty0 429--447, 2005.
\newblock ISSN 03195724.
\newblock \doi{10.1002/cjs.5540330308}.
\newblock URL \url{http://www.jstor.org/stable/25046189}.

\bibitem[Breiman et~al.(1984)Breiman, Friedman, Stone, and Olshen]{breimanClassificationRegressionTrees1984a}
Leo Breiman, Jerome Friedman, Charles~J. Stone, and R.~A. Olshen.
\newblock \emph{Classification and {{Regression Trees}}}.
\newblock {Chapman \& Hall}, 1984.
\newblock ISBN 978-0-412-04841-8.

\bibitem[Bou-Hamad et~al.(2011)Bou-Hamad, Larocque, and {Ben-Ameur}]{bou-hamadReviewSurvivalTrees2011}
Imad Bou-Hamad, Denis Larocque, and Hatem {Ben-Ameur}.
\newblock A review of survival trees.
\newblock \emph{Statistics Surveys}, 5:\penalty0 44--71, 2011.
\newblock \doi{10.1214/09-SS047}.

\bibitem[Hothorn et~al.(2006)Hothorn, Hornik, and Zeileis]{hothornUnbiasedRecursivePartitioning2006}
Torsten Hothorn, Kurt Hornik, and Achim Zeileis.
\newblock Unbiased {{Recursive Partitioning}}: {{A Conditional Inference Framework}}.
\newblock \emph{Journal of Computational and Graphical Statistics}, 15\penalty0 (3):\penalty0 651--674, 2006.
\newblock ISSN 1061-8600.
\newblock \doi{10.1198/106186006X133933}.

\bibitem[Emura et~al.(2023{\natexlab{a}})Emura, Hsu, and Chou]{emuraSurvivalTreeBased2023}
Takeshi Emura, Wei-Chern Hsu, and Wen-Chi Chou.
\newblock A survival tree based on stabilized score tests for high-dimensional covariates.
\newblock \emph{Journal of Applied Statistics}, 50\penalty0 (2):\penalty0 264--290, 2023{\natexlab{a}}.
\newblock ISSN 0266-4763.
\newblock \doi{10.1080/02664763.2021.1990224}.

\bibitem[Ciampi et~al.(1986)Ciampi, Thiffault, Nakache, and Asselain]{ciampiStratificationStepwiseRegression1986}
Antonio Ciampi, Johanne Thiffault, Jean-Pierre Nakache, and Bernard Asselain.
\newblock Stratification by stepwise regression, correspondence analysis and recursive partition: A comparison of three methods of analysis for survival data with covariates.
\newblock \emph{Computational Statistics \& Data Analysis}, 4\penalty0 (3):\penalty0 185--204, 1986.
\newblock ISSN 0167-9473.
\newblock \doi{10.1016/0167-9473(86)90033-2}.

\bibitem[Dobler and Pauly(2018)]{doblerBootstrapPermutationbasedInference2018}
Dennis Dobler and Markus Pauly.
\newblock Bootstrap- and permutation-based inference for the {{Mann}}\textendash{{Whitney}} effect for right-censored and tied data.
\newblock \emph{TEST}, 27\penalty0 (3):\penalty0 639--658, 2018.
\newblock ISSN 1863-8260.
\newblock \doi{10.1007/s11749-017-0565-z}.

\bibitem[Ditzhaus et~al.(2023{\natexlab{a}})Ditzhaus, Genuneit, Janssen, and Pauly]{ditzhausCASANOVAPermutationInference2023}
Marc Ditzhaus, Jon Genuneit, Arnold Janssen, and Markus Pauly.
\newblock {{CASANOVA}}: {{Permutation}} inference in factorial survival designs.
\newblock \emph{Biometrics}, 79\penalty0 (1):\penalty0 203--215, 2023{\natexlab{a}}.
\newblock ISSN 1541-0420.
\newblock \doi{10.1111/biom.13575}.

\bibitem[Pepe and Fleming(1989)]{pepeWeightedKaplanMeierStatistics1989}
Margaret~Sullivan Pepe and Thomas~R. Fleming.
\newblock Weighted {{Kaplan-Meier Statistics}}: {{A Class}} of {{Distance Tests}} for {{Censored Survival Data}}.
\newblock \emph{Biometrics}, 45\penalty0 (2):\penalty0 497--507, 1989.
\newblock ISSN 0006-341X.
\newblock \doi{10.2307/2531492}.

\bibitem[Moradian et~al.(2017)Moradian, Larocque, and Bellavance]{moradianSplittingRulesSurvival2017}
Hoora Moradian, Denis Larocque, and Fran{\c c}ois Bellavance.
\newblock {{L}}{$_1$} splitting rules in survival forests.
\newblock \emph{Lifetime Data Analysis}, 23\penalty0 (4):\penalty0 671--691, 2017.
\newblock ISSN 1572-9249.
\newblock \doi{10.1007/s10985-016-9372-1}.

\bibitem[Moradian et~al.(2019)Moradian, Larocque, and Bellavance]{moradianSurvivalForestsData2019}
Hoora Moradian, Denis Larocque, and Fran{\c c}ois Bellavance.
\newblock Survival forests for data with dependent censoring.
\newblock \emph{Statistical Methods in Medical Research}, 28\penalty0 (2):\penalty0 445--461, 2019.
\newblock ISSN 1477-0334.
\newblock \doi{10.1177/0962280217727314}.

\bibitem[Klein and Moeschberger(2003)]{kleinBasicQuantitiesModels2003}
John~P. Klein and Melvin~L. Moeschberger.
\newblock \emph{Survival {Analysis}: {Techniques} for {Censored} and {Truncated} {Data}}.
\newblock Springer, New York, NY, 2003.
\newblock ISBN 978-0-387-95399-1 978-0-387-21645-4.
\newblock \doi{10.1007/b97377}.
\newblock URL \url{http://link.springer.com/10.1007/b97377}.

\bibitem[Emura and Chen(2018)]{emuraCopulaModelsDependent2018}
Takeshi Emura and Yi-Hau Chen.
\newblock \emph{Analysis of Survival Data with Dependent Censoring: Copula-Based Approaches}.
\newblock {Springer}, {Singapore}, 2018.
\newblock ISBN 978-981-10-7163-8.
\newblock \doi{10.1007/978-981-10-7164-5}.

\bibitem[Rivest and Wells(2001)]{rivestMartingaleApproachCopulaGraphic2001}
Louis-Paul Rivest and Martin~T. Wells.
\newblock A {{Martingale Approach}} to the {{Copula-Graphic Estimator}} for the {{Survival Function}} under {{Dependent Censoring}}.
\newblock \emph{Journal of Multivariate Analysis}, 79\penalty0 (1):\penalty0 138--155, 2001.
\newblock ISSN 0047-259X.
\newblock \doi{10.1006/jmva.2000.1959}.

\bibitem[Lehmann and Romano(2022)]{lehmannPermutationRandomizationTests2022}
E.L. Lehmann and Joseph~P. Romano.
\newblock \emph{Testing {Statistical} {Hypotheses}}.
\newblock Springer International Publishing, Cham, 2022.
\newblock ISBN 978-3-030-70577-0 978-3-030-70578-7.
\newblock \doi{10.1007/978-3-030-70578-7}.
\newblock URL \url{https://link.springer.com/10.1007/978-3-030-70578-7}.

\bibitem[Tsiatis(1975)]{tsiatisNonidentifiabilityAspectProblem1975}
A~Tsiatis.
\newblock A nonidentifiability aspect of the problem of competing risks.
\newblock \emph{Proceedings of the National Academy of Sciences of the United States of America}, 72\penalty0 (1):\penalty0 20--22, 1975.
\newblock ISSN 0027-8424.

\bibitem[Emura and Michimae(2017)]{emuraCopulabasedInferencePiecewise2017}
Takeshi Emura and Hirofumi Michimae.
\newblock A copula-based inference to piecewise exponential models under dependent censoring, with application to time to metamorphosis of salamander larvae.
\newblock \emph{Environmental and Ecological Statistics}, 24\penalty0 (1):\penalty0 151--173, 2017.
\newblock ISSN 1573-3009.
\newblock \doi{10.1007/s10651-017-0364-4}.

\bibitem[Hsu et~al.(2016)Hsu, Emura, and Fan]{hsuReliabilityInferenceCopulaBased2016}
Tsung-Ming Hsu, Takeshi Emura, and Tsai-Hung Fan.
\newblock Reliability {{Inference}} for a {{Copula-Based Series System Life Test Under Multiple Type-I Censoring}}.
\newblock \emph{IEEE Transactions on Reliability}, 65\penalty0 (2):\penalty0 1069--1080, 2016.
\newblock ISSN 1558-1721.
\newblock \doi{10.1109/TR.2016.2515589}.

\bibitem[Bender et~al.(2005)Bender, Augustin, and Blettner]{benderGeneratingSurvivalTimes2005}
Ralf Bender, Thomas Augustin, and Maria Blettner.
\newblock Generating survival times to simulate {{Cox}} proportional hazards models.
\newblock \emph{Statistics in Medicine}, 24\penalty0 (11):\penalty0 1713--1723, June 2005.
\newblock ISSN 0277-6715.
\newblock \doi{10.1002/sim.2059}.

\bibitem[Crowther and Lambert(2013)]{crowtherSimulatingBiologicallyPlausible2013}
Michael~J. Crowther and Paul~C. Lambert.
\newblock Simulating biologically plausible complex survival data.
\newblock \emph{Statistics in Medicine}, 32\penalty0 (23):\penalty0 4118--4134, 2013.
\newblock ISSN 02776715.
\newblock \doi{10.1002/sim.5823}.

\bibitem[Dormuth et~al.(2023)Dormuth, Liu, Xu, Pauly, and Ditzhaus]{dormuthComparativeStudyAlternatives2023}
Ina Dormuth, Tiantian Liu, Jin Xu, Markus Pauly, and Marc Ditzhaus.
\newblock A comparative study to alternatives to the log-rank test.
\newblock \emph{Contemporary Clinical Trials}, 128:\penalty0 107165, 2023.
\newblock ISSN 1551-7144.
\newblock \doi{https://doi.org/10.1016/j.cct.2023.107165}.
\newblock URL \url{https://www.sciencedirect.com/science/article/pii/S1551714423000885}.

\bibitem[Wan(2017)]{wanSimulatingSurvivalData2017}
Fei Wan.
\newblock Simulating survival data with predefined censoring rates for proportional hazards models.
\newblock \emph{Statistics in Medicine}, 36\penalty0 (5):\penalty0 838--854, 2017.
\newblock ISSN 1097-0258.
\newblock \doi{10.1002/sim.7178}.

\bibitem[Emura(2022)]{emuraCopulaSurvAnalysis2022}
Takeshi Emura.
\newblock \emph{Copula.surv: Analysis of Bivariate Survival Data Based on Copulas}, 2022.
\newblock URL \url{https://CRAN.R-project.org/package=Copula.surv}.
\newblock R package version 1.2.

\bibitem[Morris et~al.(2019)Morris, White, and Crowther]{morrisUsingSimulationStudies2019}
Tim~P. Morris, Ian~R. White, and Michael~J. Crowther.
\newblock Using simulation studies to evaluate statistical methods.
\newblock \emph{Statistics in Medicine}, 38\penalty0 (11):\penalty0 2074--2102, 2019.
\newblock ISSN 1097-0258.
\newblock \doi{10.1002/sim.8086}.

\bibitem[Marozzi(2016)]{marozziMultivariateTestsBased2016}
Marco Marozzi.
\newblock Multivariate tests based on interpoint distances with application to magnetic resonance imaging.
\newblock \emph{Statistical Methods in Medical Research}, 25\penalty0 (6):\penalty0 2593--2610, 2016.
\newblock ISSN 0962-2802.
\newblock \doi{10.1177/0962280214529104}.

\bibitem[Boos and Zhang(2000)]{boosMonteCarloEvaluation2000}
Dennis~D. Boos and Ji~Zhang.
\newblock Monte {{Carlo Evaluation}} of {{Resampling-Based Hypothesis Tests}}.
\newblock \emph{Journal of the American Statistical Association}, 95\penalty0 (450):\penalty0 486--492, 2000.
\newblock ISSN 0162-1459.
\newblock \doi{10.2307/2669393}.

\bibitem[Therneau(2023)]{survival-package}
Terry~M Therneau.
\newblock \emph{A Package for Survival Analysis in {{R}}}, 2023.
\newblock URL \url{https://CRAN.R-project.org/package=survival}.
\newblock R package version 3.5-7.

\bibitem[{R Core Team}(2022)]{rcoreteamLanguageEnvironmentStatistical2022}
{R Core Team}.
\newblock \emph{R: {{A}} Language and Environment for Statistical Computing}.
\newblock {Vienna, Austria}, 2022.

\bibitem[Wickham(2016)]{wickhamGgplot2ElegantGraphics2016}
Hadley Wickham.
\newblock \emph{Ggplot2: {{Elegant}} Graphics for Data Analysis}.
\newblock {Springer-Verlag New York}, 2016.
\newblock ISBN 978-3-319-24277-4.

\bibitem[Emura et~al.(2012)Emura, Chen, and Chen]{emuraSurvivalPredictionBased2012}
Takeshi Emura, Yi-Hau Chen, and Hsuan-Yu Chen.
\newblock Survival {{Prediction Based}} on {{Compound Covariate}} under {{Cox Proportional Hazard Models}}.
\newblock \emph{PLOS ONE}, 7\penalty0 (10):\penalty0 e47627, 2012.
\newblock ISSN 1932-6203.
\newblock \doi{10.1371/journal.pone.0047627}.

\bibitem[Emura et~al.(2023{\natexlab{b}})Emura, Chen, Matsui, and Chen]{emuraCompoundCox2023}
Takeshi Emura, Hsuan-Yu Chen, Shigeyuki Matsui, and Yi-Hau Chen.
\newblock \emph{compound.Cox: Univariate Feature Selection and Compound Covariate for Predicting Survival}, 2023{\natexlab{b}}.
\newblock URL \url{https://CRAN.R-project.org/package=compound.Cox}.
\newblock R package version 3.30.

\bibitem[Emura and Hsu(2021)]{emuraUniSurvivalTree2021}
Takeshi Emura and Wei-Chern Hsu.
\newblock \emph{Uni.Survival.Tree: {{A}} Survival Tree Based on Stabilized Score Tests for High-Dimensional Covariates}, 2021.
\newblock R package version 1.5.

\bibitem[Harrell et~al.(1982)Harrell, Califf, Pryor, Lee, and Rosati]{harrellEvaluatingYieldMedical1982}
F.~E. Harrell, R.~M. Califf, D.~B. Pryor, K.~L. Lee, and R.~A. Rosati.
\newblock Evaluating the yield of medical tests.
\newblock \emph{JAMA}, 247\penalty0 (18):\penalty0 2543--2546, 1982.
\newblock ISSN 0098-7484.

\bibitem[Bertsimas et~al.(2022)Bertsimas, Dunn, Gibson, and Orfanoudaki]{bertsimasOptimalSurvivalTrees2022}
Dimitris Bertsimas, Jack Dunn, Emma Gibson, and Agni Orfanoudaki.
\newblock Optimal survival trees.
\newblock \emph{Machine Learning}, 111\penalty0 (8):\penalty0 2951--3023, 2022.
\newblock ISSN 1573-0565.
\newblock \doi{10.1007/s10994-021-06117-0}.

\bibitem[Graf et~al.(1999)Graf, Schmoor, Sauerbrei, and Schumacher]{grafAssessmentComparisonPrognostic1999}
Erika Graf, Claudia Schmoor, Willi Sauerbrei, and Martin Schumacher.
\newblock Assessment and comparison of prognostic classification schemes for survival data.
\newblock \emph{Statistics in Medicine}, 18\penalty0 (17-18):\penalty0 2529--2545, 1999.
\newblock ISSN 1097-0258.
\newblock \doi{10.1002/(SICI)1097-0258(19990915/30)18:17/18<2529::AID-SIM274>3.0.CO;2-5}.

\bibitem[Zhou et~al.(2022)Zhou, Cheng, Wang, Zou, and Wang]{zhouSurvMetricsPredictiveEvaluation2022}
Hanpu Zhou, Xuewei Cheng, Sizheng Wang, Yi~Zou, and Hong Wang.
\newblock \emph{{{SurvMetrics}}: {{Predictive}} Evaluation Metrics in Survival Analysis}, 2022.
\newblock R package version 0.5.0.

\bibitem[Therneau and Grambsch(2000)]{therneauCoxModel2000}
Terry~M. Therneau and Patricia~M. Grambsch.
\newblock \emph{Modeling {Survival} {Data}: {Extending} the {Cox} {Model}}.
\newblock Springer, New York, NY, 2000.
\newblock ISBN 978-1-4419-3161-0 978-1-4757-3294-8.
\newblock \doi{10.1007/978-1-4757-3294-8}.
\newblock URL \url{http://link.springer.com/10.1007/978-1-4757-3294-8}.

\bibitem[Hastie et~al.(2009)Hastie, Tibshirani, and Friedman]{hastieModelAssessmentSelection2009}
Trevor Hastie, Robert Tibshirani, and Jerome Friedman.
\newblock \emph{The {Elements} of {Statistical} {Learning}}.
\newblock Springer, New York, NY, 2009.
\newblock ISBN 978-0-387-84857-0 978-0-387-84858-7.
\newblock \doi{10.1007/978-0-387-84858-7}.
\newblock URL \url{http://link.springer.com/10.1007/978-0-387-84858-7}.

\bibitem[Ciampi et~al.(1995)Ciampi, Negassa, and Lou]{ciampiTreestructuredPredictionCensored1995}
A.~Ciampi, A.~Negassa, and Z.~Lou.
\newblock Tree-structured prediction for censored survival data and the {{Cox}} model.
\newblock \emph{Journal of Clinical Epidemiology}, 48\penalty0 (5):\penalty0 675--689, 1995.
\newblock ISSN 0895-4356.
\newblock \doi{10.1016/0895-4356(94)00164-l}.

\bibitem[Brendel et~al.(2014)Brendel, Janssen, Mayer, and Pauly]{brendelWeightedLogrankPermutation2014}
Michael Brendel, Arnold Janssen, Claus-Dieter Mayer, and Markus Pauly.
\newblock Weighted {{Logrank Permutation Tests}} for {{Randomly Right Censored Life Science Data}}.
\newblock \emph{Scandinavian Journal of Statistics}, 41\penalty0 (3):\penalty0 742--761, 2014.
\newblock ISSN 0303-6898.
\newblock \doi{10.2307/24586756}.

\bibitem[Ditzhaus and Friedrich(2020)]{ditzhausMorePowerfulLogrank2020}
Marc Ditzhaus and Sarah Friedrich.
\newblock More powerful logrank permutation tests for two-sample survival data.
\newblock \emph{Journal of Statistical Computation and Simulation}, 90\penalty0 (12):\penalty0 2209--2227, 2020.
\newblock ISSN 0094-9655.
\newblock \doi{10.1080/00949655.2020.1773463}.

\bibitem[Ditzhaus et~al.(2023{\natexlab{b}})Ditzhaus, Yu, and Xu]{ditzhausStudentizedPermutationMethod2023}
Marc Ditzhaus, Menggang Yu, and Jin Xu.
\newblock Studentized permutation method for comparing two restricted mean survival times with small sample from randomized trials.
\newblock \emph{Statistics in Medicine}, 42\penalty0 (13):\penalty0 2226--2240, 2023{\natexlab{b}}.
\newblock ISSN 1097-0258.
\newblock \doi{10.1002/sim.9720}.

\bibitem[Ishwaran et~al.(2008)Ishwaran, Kogalur, Blackstone, and Lauer]{ishwaran}
Hemant Ishwaran, Udaya~B. Kogalur, Eugene~H. Blackstone, and Michael~S. Lauer.
\newblock {Random survival forests}.
\newblock \emph{The Annals of Applied Statistics}, 2\penalty0 (3):\penalty0 841 -- 860, 2008.
\newblock \doi{10.1214/08-AOAS169}.
\newblock URL \url{https://doi.org/10.1214/08-AOAS169}.

\end{thebibliography}

\newpage
\appendix

\section{Tables\label{app2}}%
\vspace*{12pt}

\begin{table}[ht]
\centering
\caption{Type I error rate estimates (Monte Carlo standard errors) for $n_1 = n_2  =50$ by assumed dependency parameter $\tau_{theor.}$ on Clayton copula generated data.} \label{tab:type1n50}
\begin{tabular}{llllp{0.001cm}llp{0.001cm}llp{0.001cm}llp{0.001cm}ll}
\hline
 $\tau_{theor.}$ &cens.  \%  & \multicolumn{11}{c}{copula-graphic estimator permutation test} && \multicolumn{2}{c}{logrank test}\\
&&  \multicolumn{2}{c}{$\tau_{assum.}=0$}  & & \multicolumn{2}{c}{$\tau_{assum.}=0.25$} && \multicolumn{2}{c}{$\tau_{assum.}=0.5$} && \multicolumn{2}{c}{$\tau_{assum.}=0.75$} &&     \\ 
\hline
&0.100 & 0.044 & (0.005) && 0.043 & (0.005) && 0.040 & (0.004) && 0.043 & (0.005) && 0.048 & (0.005) \\ 
0&0.250 & 0.051 & (0.005) && 0.054 & (0.005) && 0.056 & (0.005) && 0.052 & (0.005) && 0.053 & (0.005) \\ 
&0.500 & 0.041 & (0.004) && 0.047 & (0.005) && 0.047 & (0.005) && 0.052 & (0.005) && 0.046 & (0.005) \\ \hline
&0.066 & 0.048 & (0.005) && 0.048 & (0.005) && 0.051 & (0.005) && 0.048 & (0.005) && 0.054 & (0.005) \\ 
0.25&0.192 & 0.050 & (0.005) && 0.046 & (0.005) && 0.048 & (0.005) && 0.046 & (0.005) && 0.056 & (0.005) \\ 
&0.501 & 0.051 & (0.005) && 0.047 & (0.005) && 0.044 & (0.005) && 0.046 & (0.005) && 0.055 & (0.005) \\ \hline
&0.038 & 0.048 & (0.005) && 0.048 & (0.005) && 0.051 & (0.005) && 0.052 & (0.005) && 0.054 & (0.005) \\ 
0.5&0.123 & 0.054 & (0.005) && 0.055 & (0.005) && 0.055 & (0.005) && 0.054 & (0.005) && 0.055 & (0.005) \\ 
&0.501 & 0.050 & (0.005) && 0.052 & (0.005) && 0.047 & (0.005) && 0.047 & (0.005) && 0.049 & (0.005) \\ \hline
&0.017 & 0.049 & (0.005) && 0.051 & (0.005) && 0.051 & (0.005) && 0.048 & (0.005) && 0.048 & (0.005) \\ 
0.75&0.060 & 0.048 & (0.005) && 0.051 & (0.005) && 0.052 & (0.005) && 0.054 & (0.005) && 0.055 & (0.005) \\ 
&0.499 & 0.051 & (0.005) && 0.048 & (0.005) && 0.048 & (0.005) && 0.048 & (0.005) && 0.052 & (0.005) \\ 
\hline
\end{tabular}
\end{table}

\begin{table}[h]
\centering
\caption{Type I error rate estimates (Monte Carlo standard errors) for $n_1 = n_2  =20$ by assumed dependency parameter $\tau_{theor.}$ on Clayton copula generated data.} \label{tab:type1n20}
\begin{tabular}{llllp{0.001cm}llp{0.001cm}llp{0.001cm}llp{0.001cm}ll}
\hline
 $\tau_{theor.}$ &cens.  \%  & \multicolumn{11}{c}{copula-graphic estimator permutation test} && \multicolumn{2}{c}{logrank test}\\
&&  \multicolumn{2}{c}{$\tau_{assum.}=0$}  & & \multicolumn{2}{c}{$\tau_{assum.}=0.25$} && \multicolumn{2}{c}{$\tau_{assum.}=0.5$} && \multicolumn{2}{c}{$\tau_{assum.}=0.75$} &&     \\ 
\hline
&0.100 & 0.040 & (0.004) && 0.043 & (0.005) && 0.046 & (0.005) && 0.041 & (0.004) && 0.054 & (0.005) \\ 
0&0.247 & 0.054 & (0.005) && 0.056 & (0.005) && 0.055 & (0.005) && 0.056 & (0.005) && 0.057 & (0.005) \\ 
&0.499 & 0.050 & (0.005) && 0.046 & (0.005) && 0.048 & (0.005) && 0.046 & (0.005) && 0.059 & (0.005) \\ \hline
&0.065 & 0.052 & (0.005) && 0.046 & (0.005) && 0.046 & (0.005) && 0.046 & (0.005) && 0.057 & (0.005) \\ 
0.25&0.192 & 0.053 & (0.005) && 0.046 & (0.005) && 0.046 & (0.005) && 0.046 & (0.005) && 0.060 & (0.005) \\ 
&0.498 & 0.051 & (0.005) && 0.050 & (0.005) && 0.046 & (0.005) && 0.051 & (0.005) && 0.052 & (0.005) \\ \hline
&0.038 & 0.056 & (0.005) && 0.054 & (0.005) && 0.054 & (0.005) && 0.053 & (0.005) && 0.064 & (0.005) \\ 
0.5&0.126 & 0.048 & (0.005) && 0.045 & (0.005) && 0.048 & (0.005) && 0.049 & (0.005) && 0.052 & (0.005) \\ 
&0.502 & 0.058 & (0.005) && 0.060 & (0.005) && 0.058 & (0.005) && 0.059 & (0.005) && 0.058 & (0.005) \\ \hline
&0.017 & 0.061 & (0.005) && 0.060 & (0.005) && 0.060 & (0.005) && 0.060 & (0.005) && 0.067 & (0.006) \\ 
0.75&0.059 & 0.051 & (0.005) && 0.051 & (0.005) && 0.054 & (0.005) && 0.052 & (0.005) && 0.062 & (0.005) \\ 
&0.500 & 0.048 & (0.005) && 0.050 & (0.005) && 0.053 & (0.005)& & 0.052 & (0.005) && 0.054 & (0.005) \\ 

\hline
\end{tabular}
\end{table}

\begin{table}[h]
\centering
\caption{Type I error rate estimates(Monte Carlo standard errors) for $n_1 = n_2  =200$ by assumed dependency parameter $\tau_{theor.}$ on Clayton copula generated data.} \label{tab:type1n200}
\begin{tabular}{llllp{0.001cm}llp{0.001cm}llp{0.001cm}llp{0.001cm}ll}
\hline
 $\tau_{theor.}$ &cens.  \%  & \multicolumn{11}{c}{copula-graphic estimator permutation test} && \multicolumn{2}{c}{logrank test}\\
&&  \multicolumn{2}{c}{$\tau_{assum.}=0$}  & & \multicolumn{2}{c}{$\tau_{assum.}=0.25$} && \multicolumn{2}{c}{$\tau_{assum.}=0.5$} && \multicolumn{2}{c}{$\tau_{assum.}=0.75$} &&     \\ 
\hline
&0.100 & 0.051 & (0.005) && 0.045 & (0.005) && 0.047 & (0.005) && 0.049 & (0.005) && 0.047 & (0.005) \\ 
0&0.250 & 0.043 & (0.005) && 0.043 & (0.005) && 0.046 & (0.005) && 0.046 & (0.005) && 0.050 & (0.005) \\ 
&0.500 & 0.048 & (0.005) && 0.052 & (0.005) && 0.054 & (0.005) && 0.058 & (0.005) && 0.053 & (0.005) \\ \hline
&0.066 & 0.047 & (0.005) && 0.048 & (0.005) && 0.049 & (0.005) && 0.050 & (0.005) && 0.048 & (0.005) \\ 
0.25&0.191 & 0.056 & (0.005) && 0.057 & (0.005) && 0.056 & (0.005) && 0.056 & (0.005) && 0.059 & (0.005) \\ 
&0.501 & 0.052 & (0.005) && 0.057 & (0.005) && 0.056 & (0.005) && 0.056 & (0.005) && 0.050 & (0.005) \\ \hline
&0.039 & 0.057 & (0.005) && 0.056 & (0.005) && 0.054 & (0.005) && 0.057 & (0.005) && 0.059 & (0.005) \\ 
0.5&0.124 & 0.043 & (0.005) && 0.047 & (0.005) && 0.046 & (0.005) && 0.046 & (0.005) && 0.047 & (0.005) \\ 
&0.500 & 0.049 & (0.005) && 0.045 & (0.005) && 0.046 & (0.005) && 0.041 & (0.004) && 0.049 & (0.005) \\ \hline
&0.017 & 0.052 & (0.005) && 0.054 & (0.005) && 0.053 & (0.005) && 0.055 & (0.005) && 0.049 & (0.005) \\ 
0.75&0.060 & 0.053 & (0.005) && 0.052 & (0.005) && 0.052 & (0.005) && 0.055 & (0.005) && 0.057 & (0.005) \\ 
&0.500 & 0.049 & (0.005) && 0.047 & (0.005) && 0.052 & (0.005) && 0.056 & (0.005) && 0.051 & (0.005) \\ 
\hline
\end{tabular}
\end{table}

\begin{table}[h]
\centering
\caption{Exemplary mean censoring proportions of $n_{sim} = 1000$ datasets for $n_1=n_2=150$, $\tau_{theor.}=0.25$ for the Clayton copula and $r=0.5$ by $\beta$ and group.} \label{tab:censoring.power}
\begin{tabular}{rrrrrrrrrrrr}
  \hline
  \hline
$\beta$ & -1.400 & -1.000 & -0.600 & -0.400 & -0.200 & 0.000 & 0.200 & 0.400 & 0.600 & 1.000 & 1.400 \\ \hline
  Group 1 & 0.502 & 0.498 & 0.498 & 0.499 & 0.498 & 0.500 & 0.501 & 0.501 & 0.500 & 0.501 & 0.499 \\ 
  Group 2 & 0.657 & 0.646 & 0.594 & 0.560 & 0.532 & 0.496 & 0.465 & 0.435 & 0.412 & 0.337 & 0.322 \\ 
   \hline
\end{tabular}
\end{table}

\vspace*{15cm}
\section{Figures}
\vspace*{12pt}

\begin{figure}[h]
\centering
\includegraphics[scale=0.505]{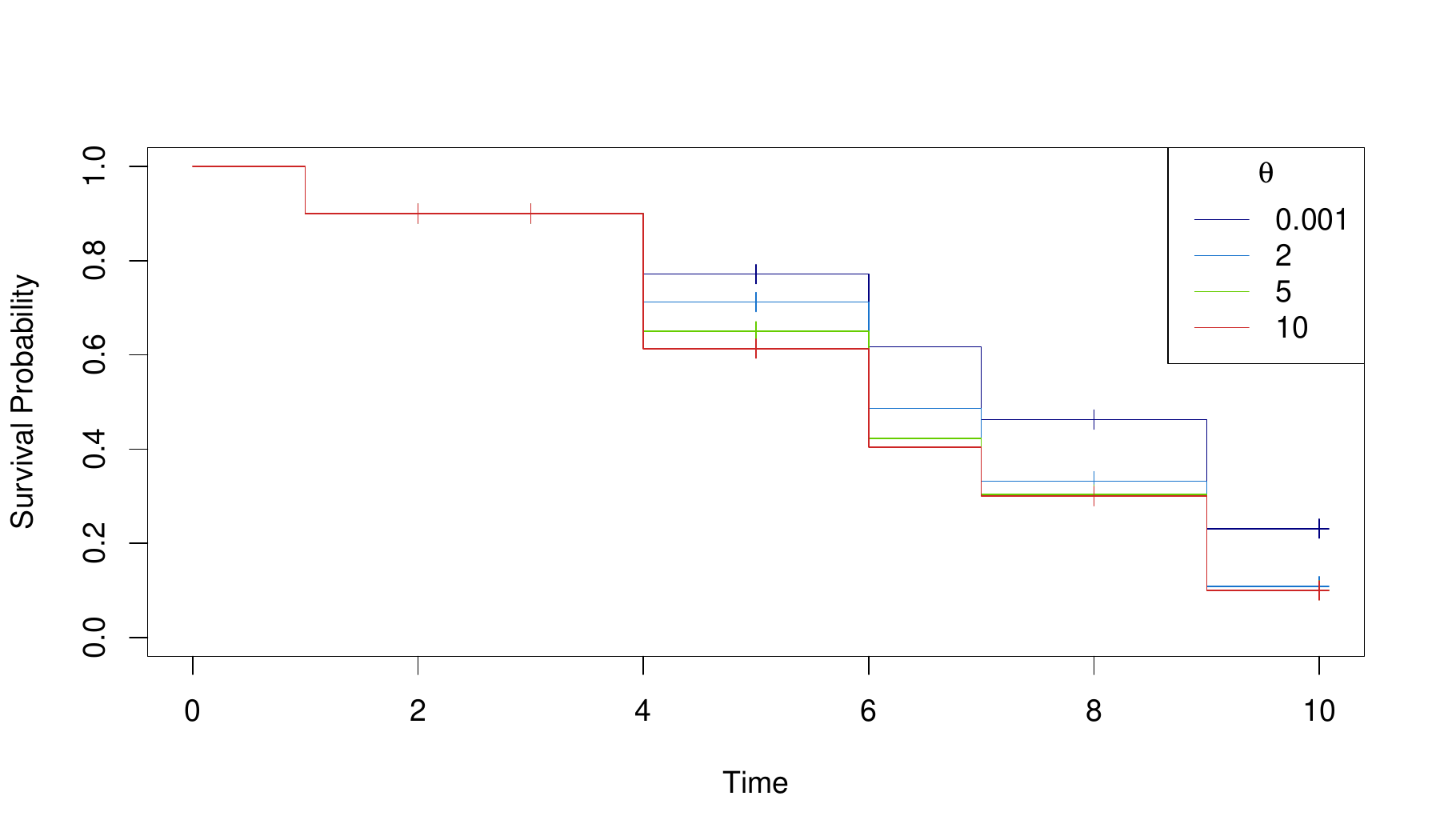}
\caption{  Copula-graphic estimator with Clayton copula and varying dependency parameter $\theta$ for data $\boldsymbol{x} = (1,2,3,4,5,6,7,8,9,10)^\top$ and $\boldsymbol{\delta} = (1,0,0,1,0,1,1,0,1,1)^\top$. Censorings are marked by an $+$-symbol. See, how increasing $\theta$ assumes a stronger dependency of event and censoring times and thus leads to a stronger influence of censorings on the jump size of the copula-graphic estimator. \label{clayton2}}
\end{figure}

\begin{figure}[h]
\centering
\includegraphics[scale=0.75]{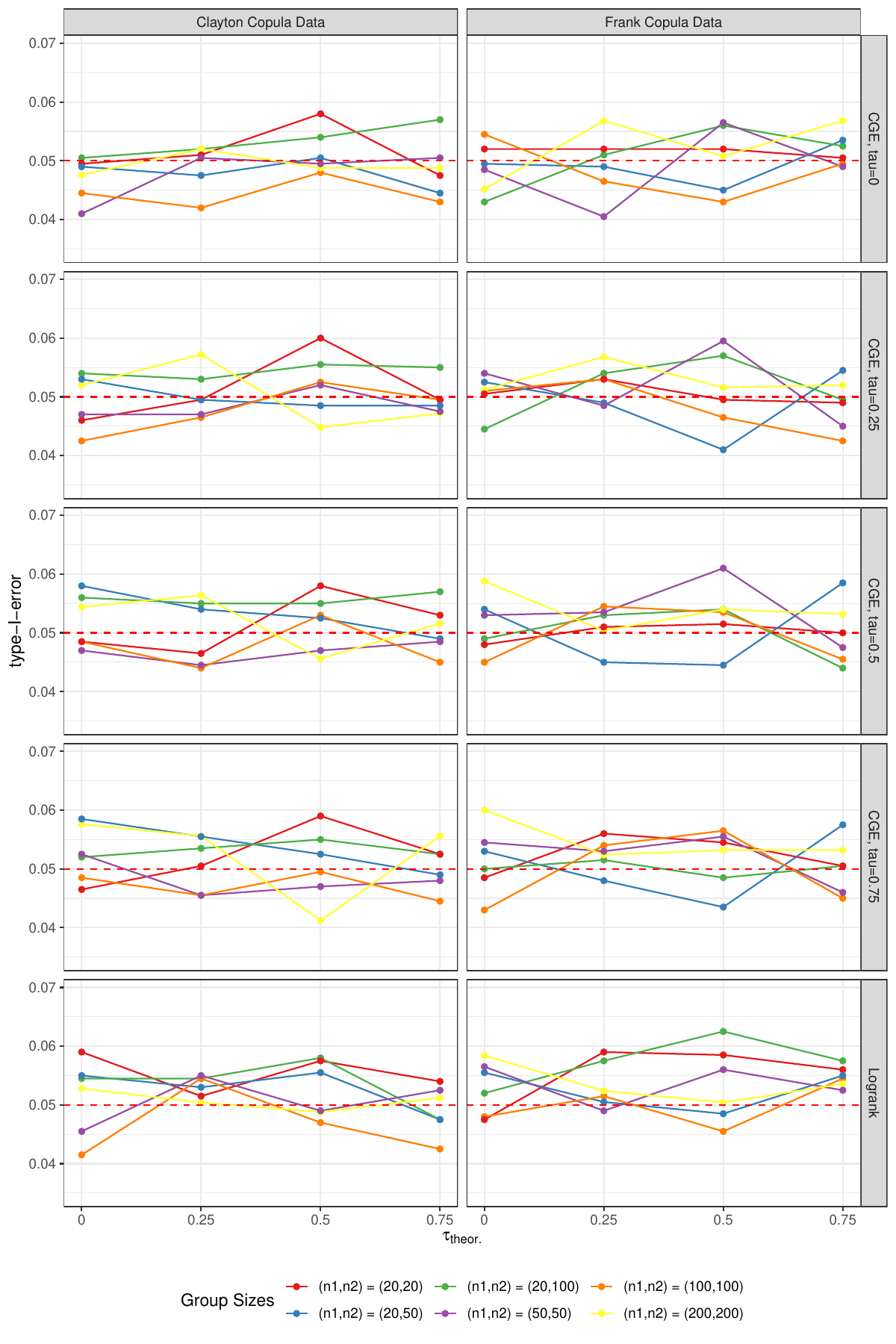}
\caption{Type I error estimates by copula model, sample size and test. Each row of the graphic displays the data of one statistical test for $r=0.5$.  \label{fig:type1error.by.copula}}
\end{figure}

\begin{figure}[h]
\centering
\includegraphics[scale=0.58]{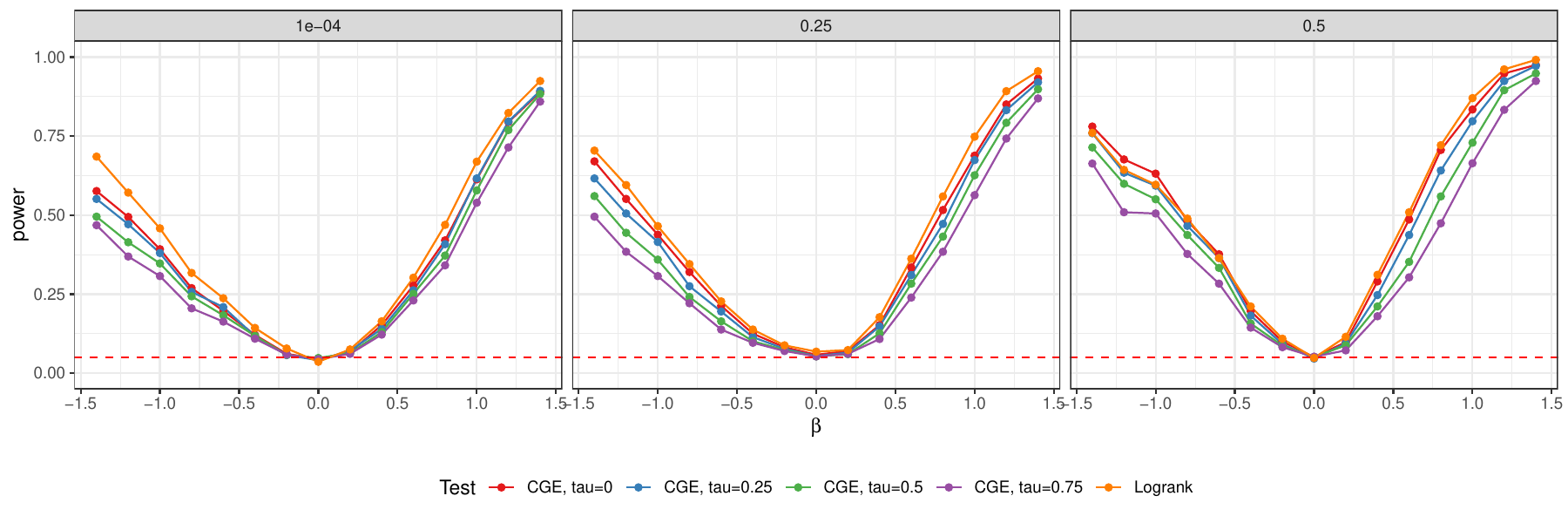}
\caption{Power estimates with theoretical dependency of event an censoring times of $n_1=n_2=20$, $r=0.5$, $\tau_{theor.} = 0.0001$ (left), $\tau_{theor.} = 0.25$ (middle) and  $\tau_{theor.} = 0.5$ (right). \label{plot:power.small.n}}
\end{figure}

\begin{figure}[h]
\centering
\includegraphics[scale=0.4]{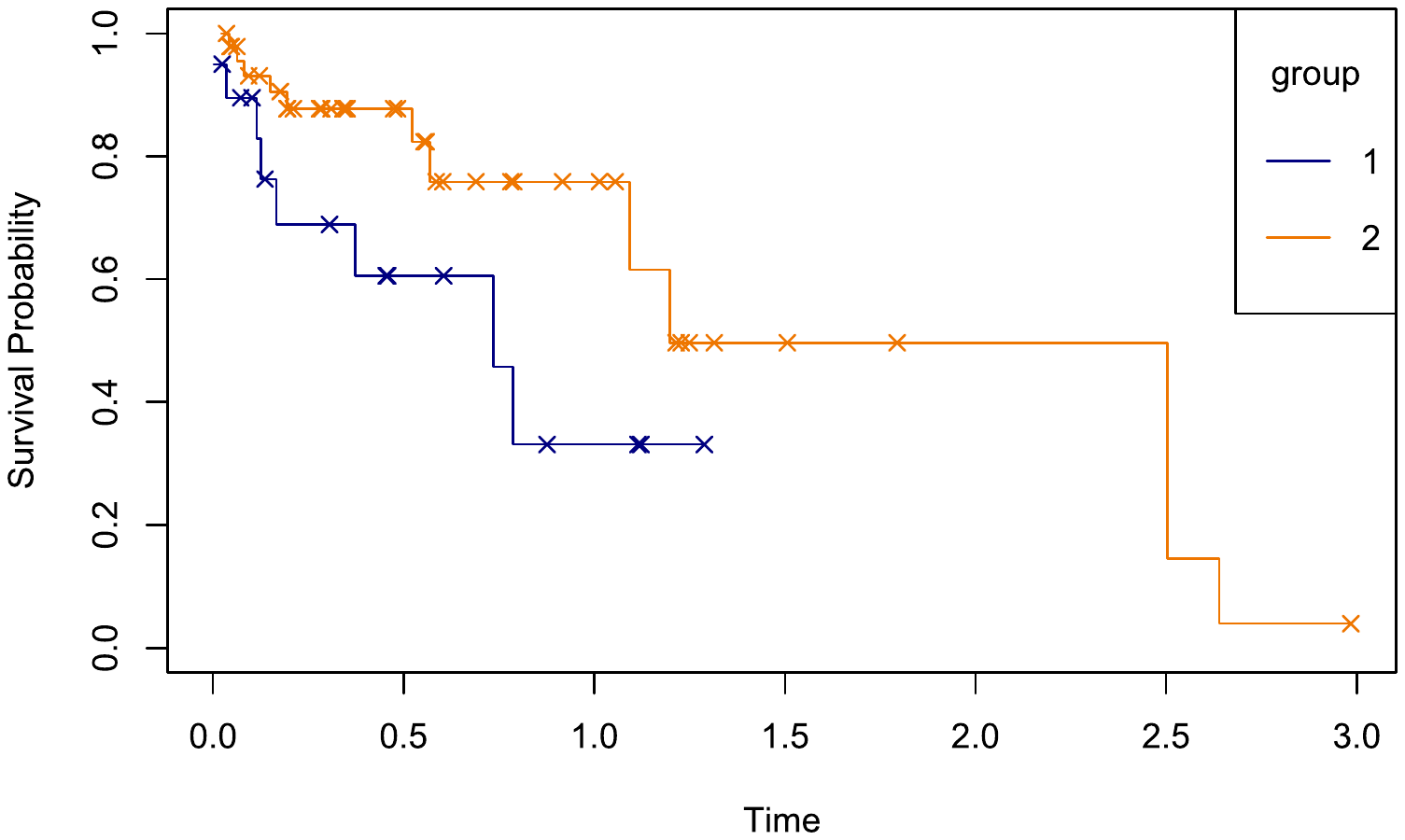}
\includegraphics[scale=0.4]{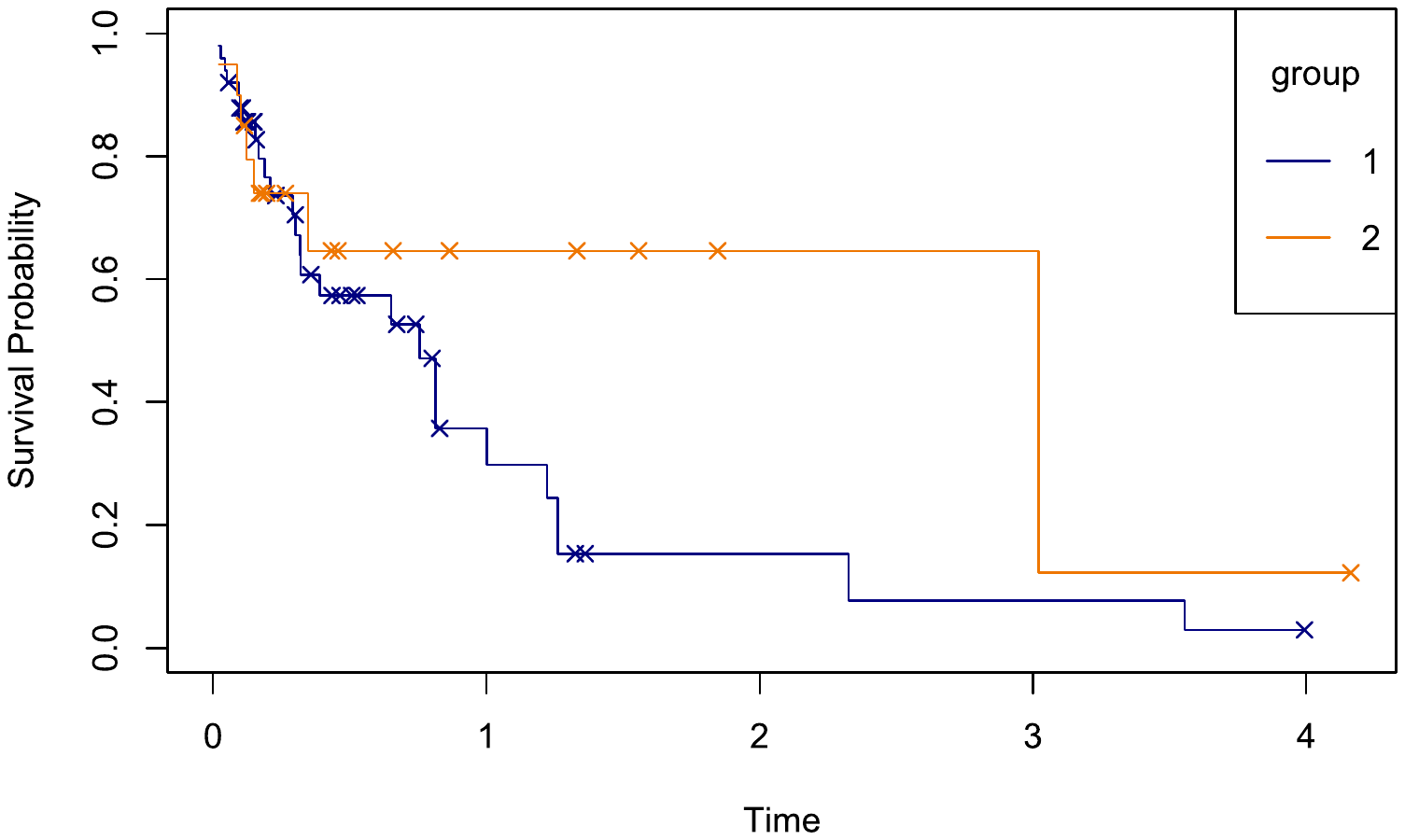}
\caption{Copula-graphic estimators with $\tau=0.25$ and the Clayton copula for exemplary data simulated with $\beta = -0.6$, $\tau_{theor.}=0.25$ in the Clayton copula, $r=0.5$ and binary covariates. $n_1=20$ and $n_2=50$ (top) and $n_1=50$ and $n_2=20$  (bottom).\newline $p$-values top: 0.170 (CGE test with $\tau_{theor.}=0.25$) and 0.031 (logrank test). $p$-values bottom: 0.030 (CGE test with $\tau_{theor.}=0.25$) and 0.141 (logrank test). }

\label{plot:cge.ex.curve}
\end{figure}

\begin{figure}
\centering
\includegraphics[scale=0.58]{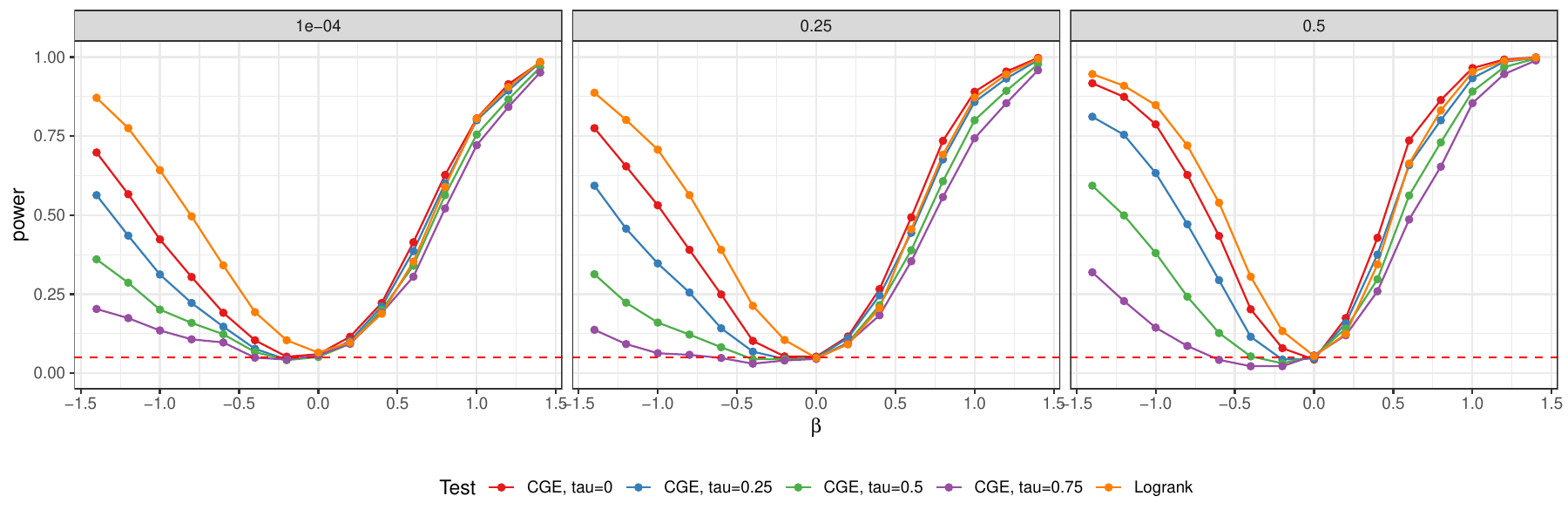}
\caption{Power estimates for unbalanced sample sizes by $\tau_{theor.}$ for $n_1=20$, $n_2=50$ and $r=0.5$. \label{plot:power.unbalanced.bytau}}
\end{figure}

\begin{figure}[h]
\centering
\includegraphics[scale=0.58]{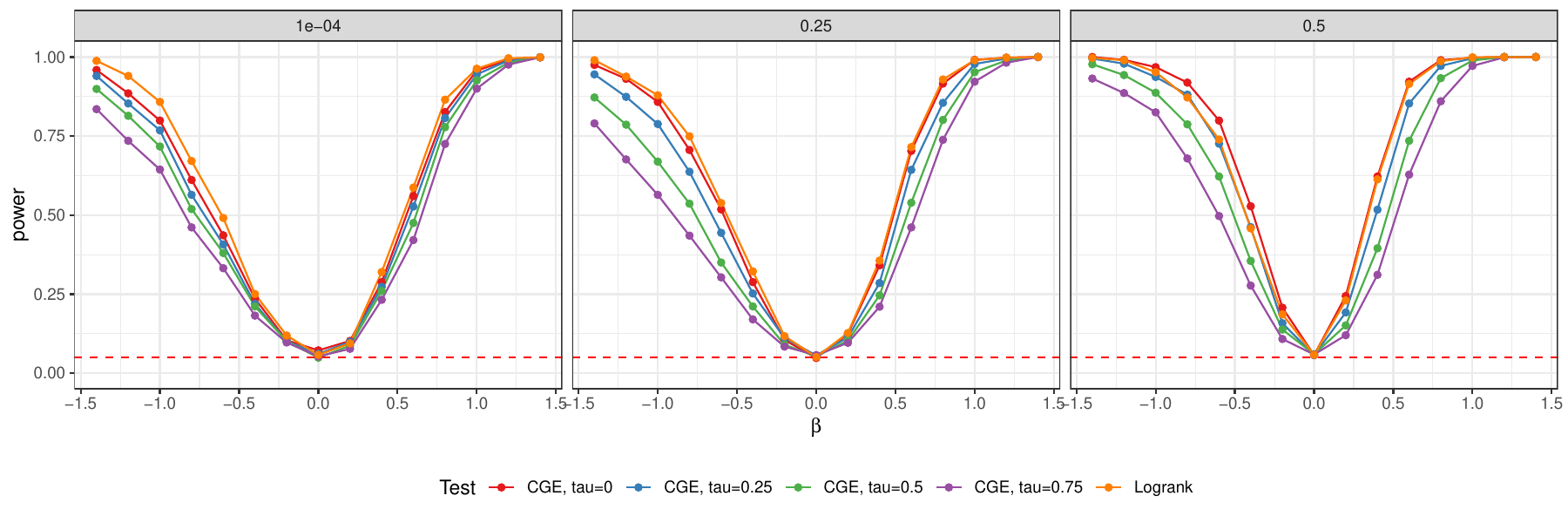}
\caption{Power estimates with theoretical dependency of event an censoring times of $n_1=n_2=50$, $r=0.5$, $\tau_{theor.} = 0.0001$ (left), $\tau_{theor.} = 0.25$ (middle) and  $\tau_{theor.} = 0.5$ (right). \label{plot:power.medium.n}}
\end{figure}

\begin{figure}
\centering
\includegraphics[scale=0.58]{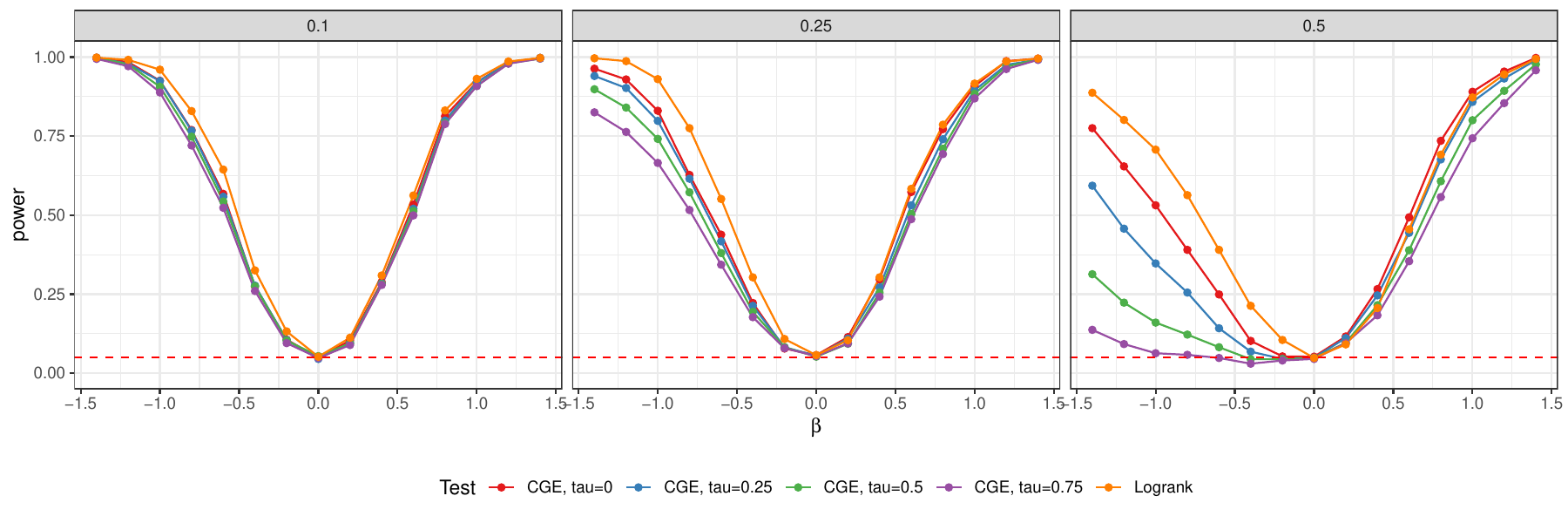}
\caption{Power estimates for unbalanced sample sizes by censoring parameter $r$ for $n_1=20$, $n_2=50$ and $\tau_{theor.}=0.25$. \label{plot:power.unbalanced.byr}}
\end{figure}

\begin{figure}
\centering
\includegraphics[scale=0.68]{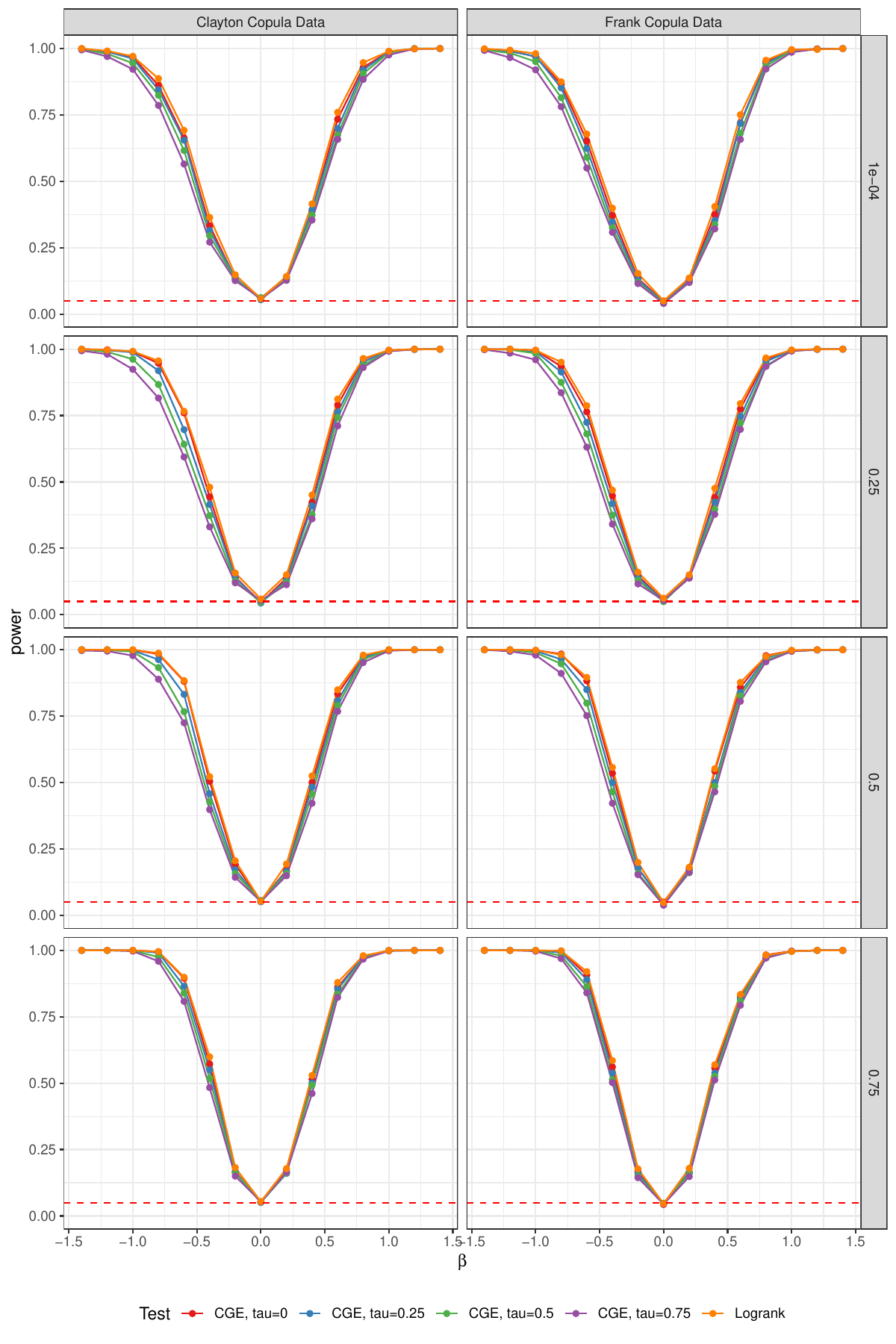}
\caption{Power estimates for $n_1=n_2=50$, $r=0.25$ and binary covariates. Data was generated using a Calyton copula (left) or a Frank copula (right). $\tau_{theor.}$ varyies by row. \label{plot:power.copula}}
\end{figure}

\begin{figure}[h]
\centering
\includegraphics[scale=0.68]{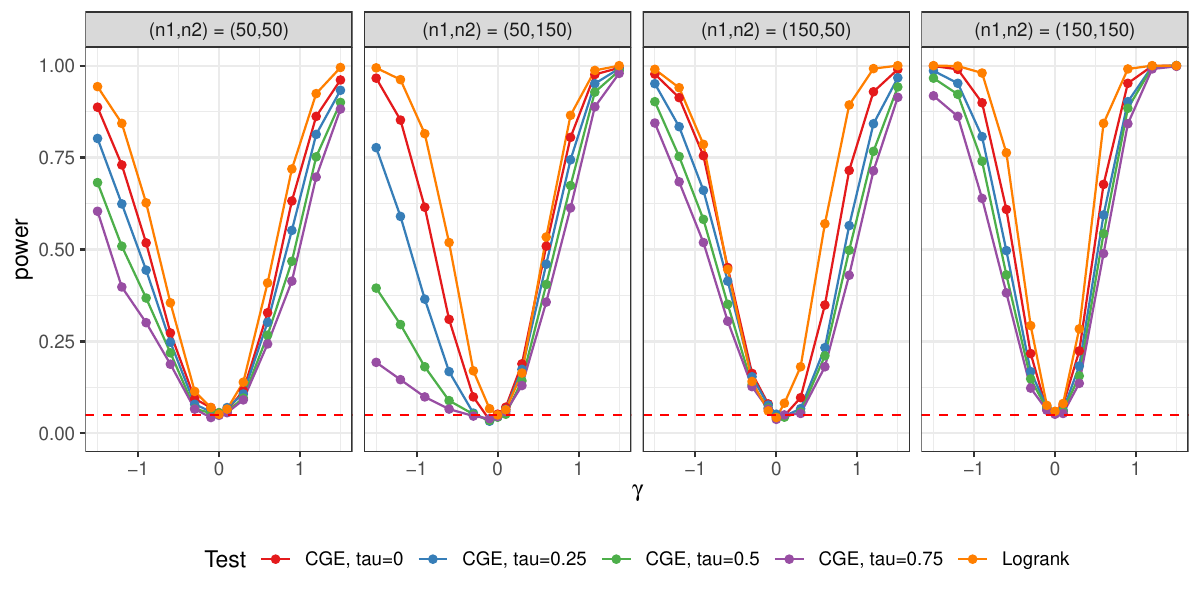}
\caption{Power estimates for normal covariables with varying mean between groups. $\tau_{theor.}=0.25$ and $r=0.5$. \label{plot:power.unbalanced.normal}}
\end{figure}

\begin{figure}[h]
\centering
\includegraphics[scale=0.68]{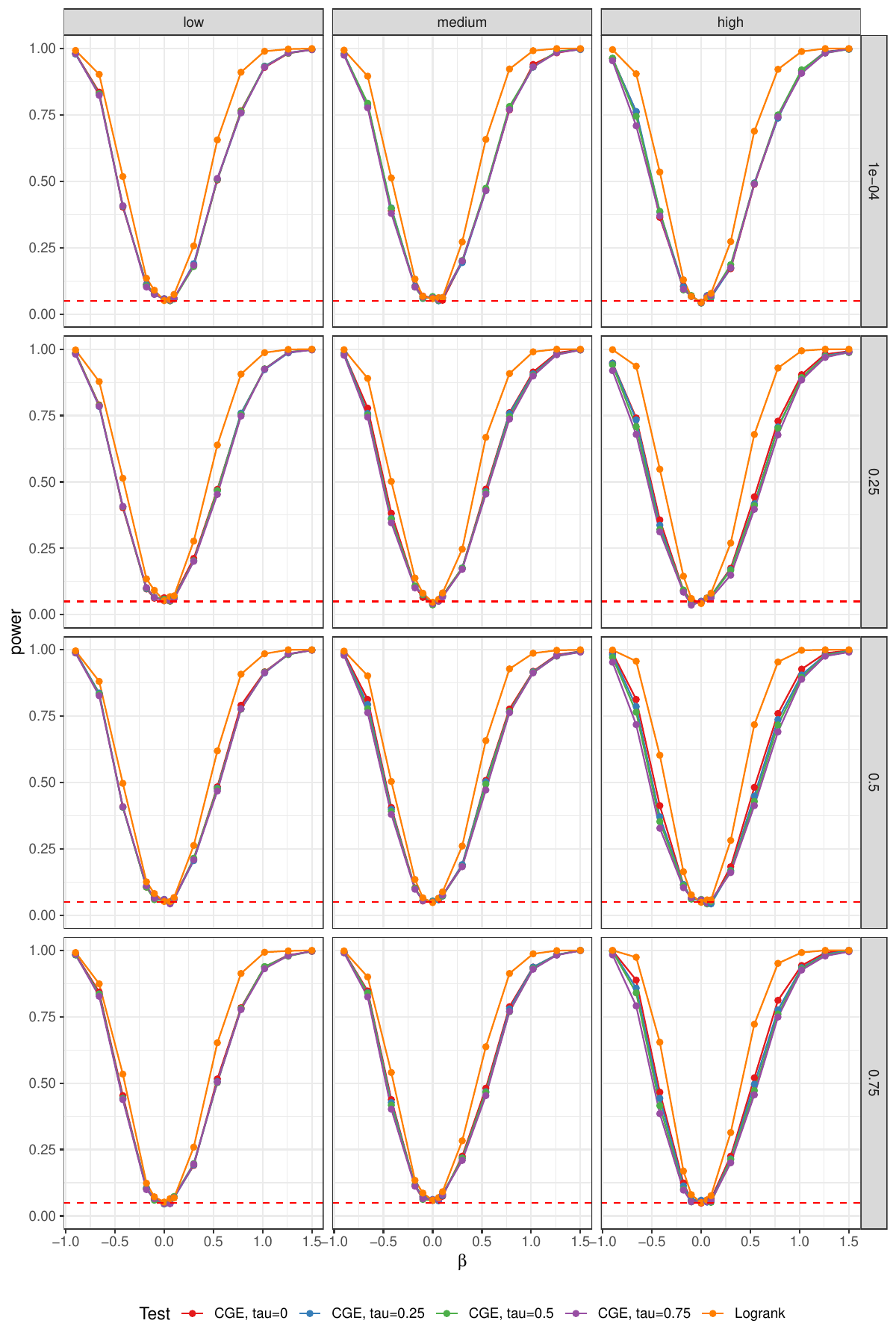}
\caption{Power estimates for poisson covariates with $n_1=n_2=150$ by censoring scenarios $r=0.1$ (low), $r=0.25$ (medium) and $r=0.5$ (high). $\tau_{theor.}$ varies across rows.  \label{plot:power.poisson}}
\end{figure}

\begin{figure}
\centering
\includegraphics[scale=0.58]{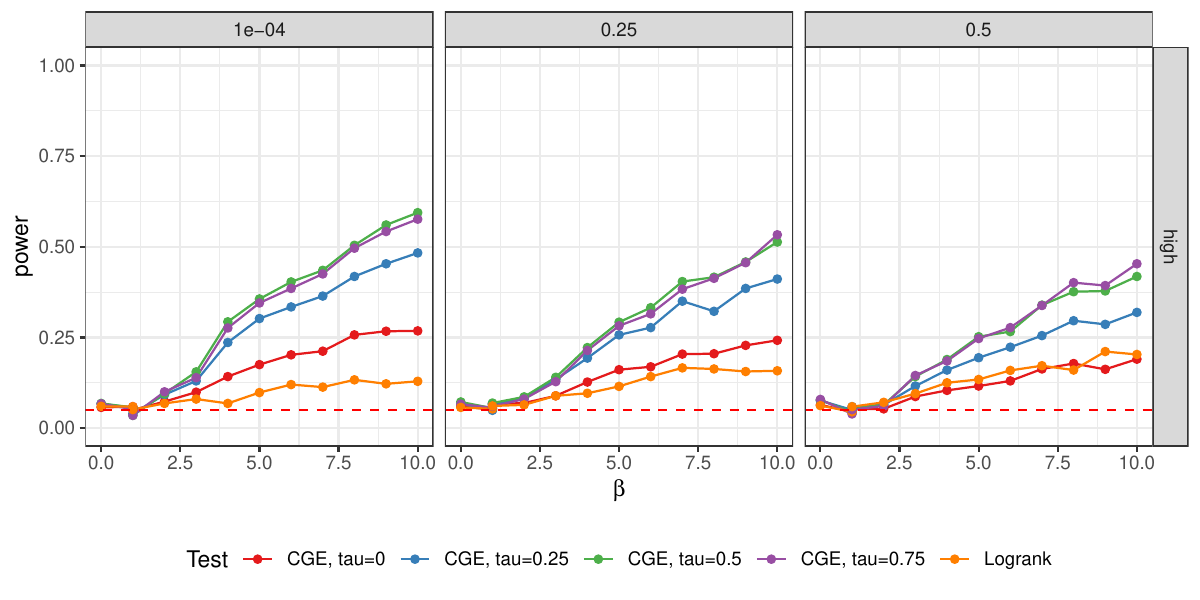}
\caption{Power estimates for normal covariates with varying standard deviation between groups, $n_1=n_2=50$ and $r=0.5$. $\tau_{theor.}$ varies across columns. \label{plot:power.normal.var.tau}}
\end{figure}

\begin{figure}
\centering
\includegraphics[scale=0.75]{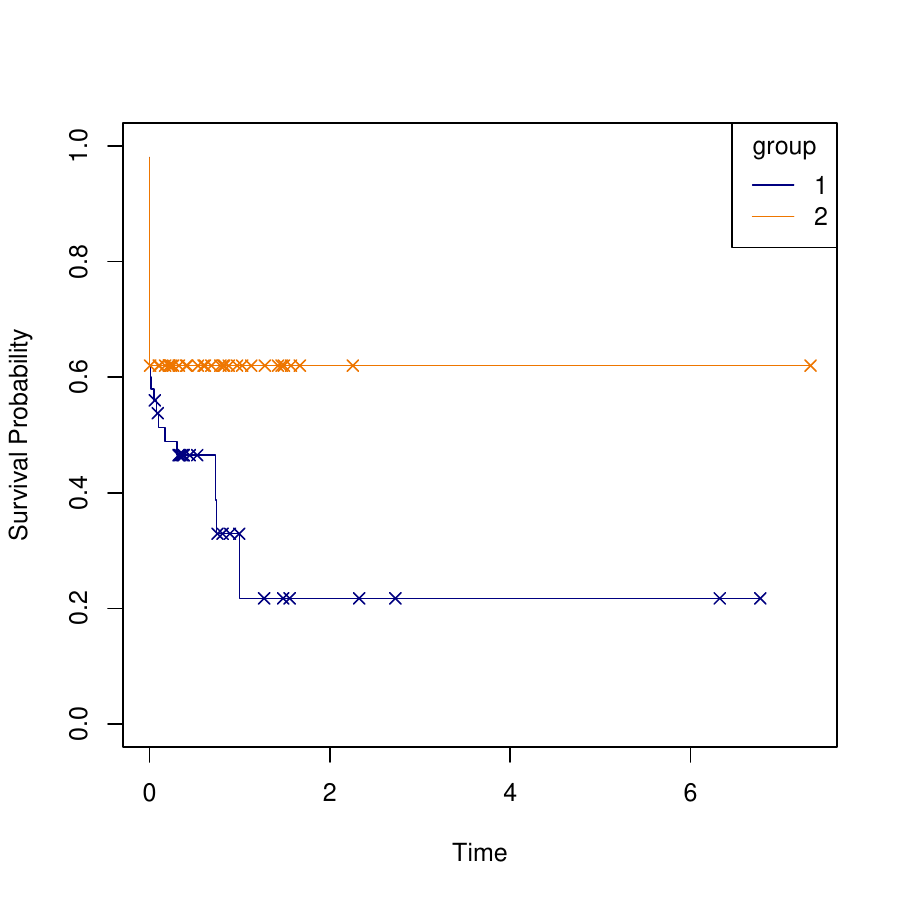}
\caption{Copula-graphic estimators with $\tau=0.25$ and the Clayton copula for exemplary data simulated with normal covariates with mean 0, stanard deviation 1 in group 1 and standard deviation $\gamma = 10$ in group 2. $\tau_{theor.}=0.5$ in the Clayton copula, $r=0.25$ and  $n_1=n_2=50$. 
$p$-values: 0.032 (CGE test with $\tau_{theor.}=0.5$) and 0.236 (logrank test).}

\label{plot:cge.ex.curve.var.normal}
\end{figure}

\begin{figure}
\centering
\includegraphics[scale=0.5]{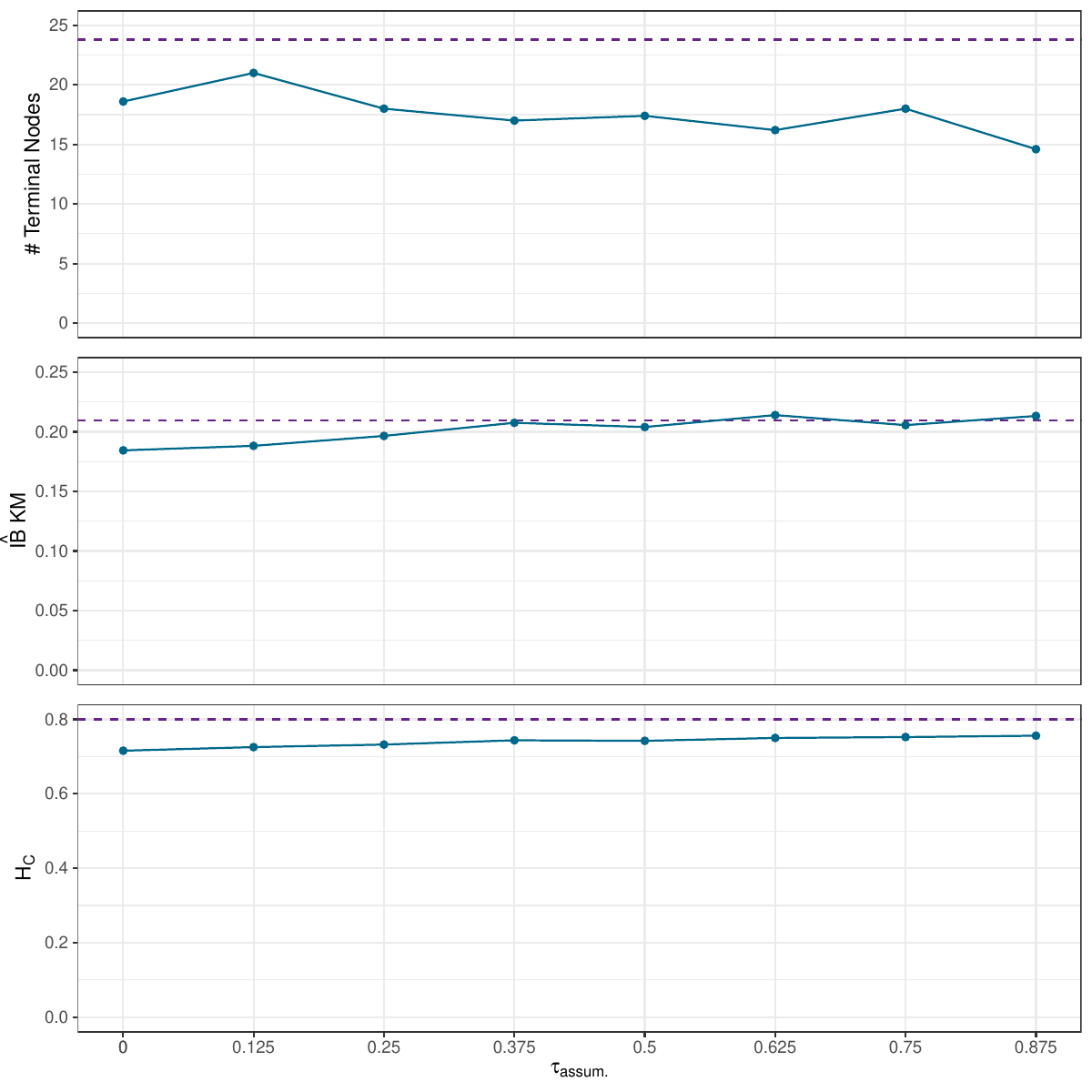}
\caption{Mean performance measures on PBC data over 5-fold cross validation. $H_C$ denotes Harrell's $C$-index and $\widehat{IB}$ KM is the Integrated Brier Score based on the Kaplan-Meier estimator. The solid lines display the data of the CGE trees by $\tau_{assum.}$ and the dashed lines display the results of the logrank tree.  \label{fig:ex.measure}}
\end{figure}

\begin{figure}
\centering
\includegraphics[scale=0.342]{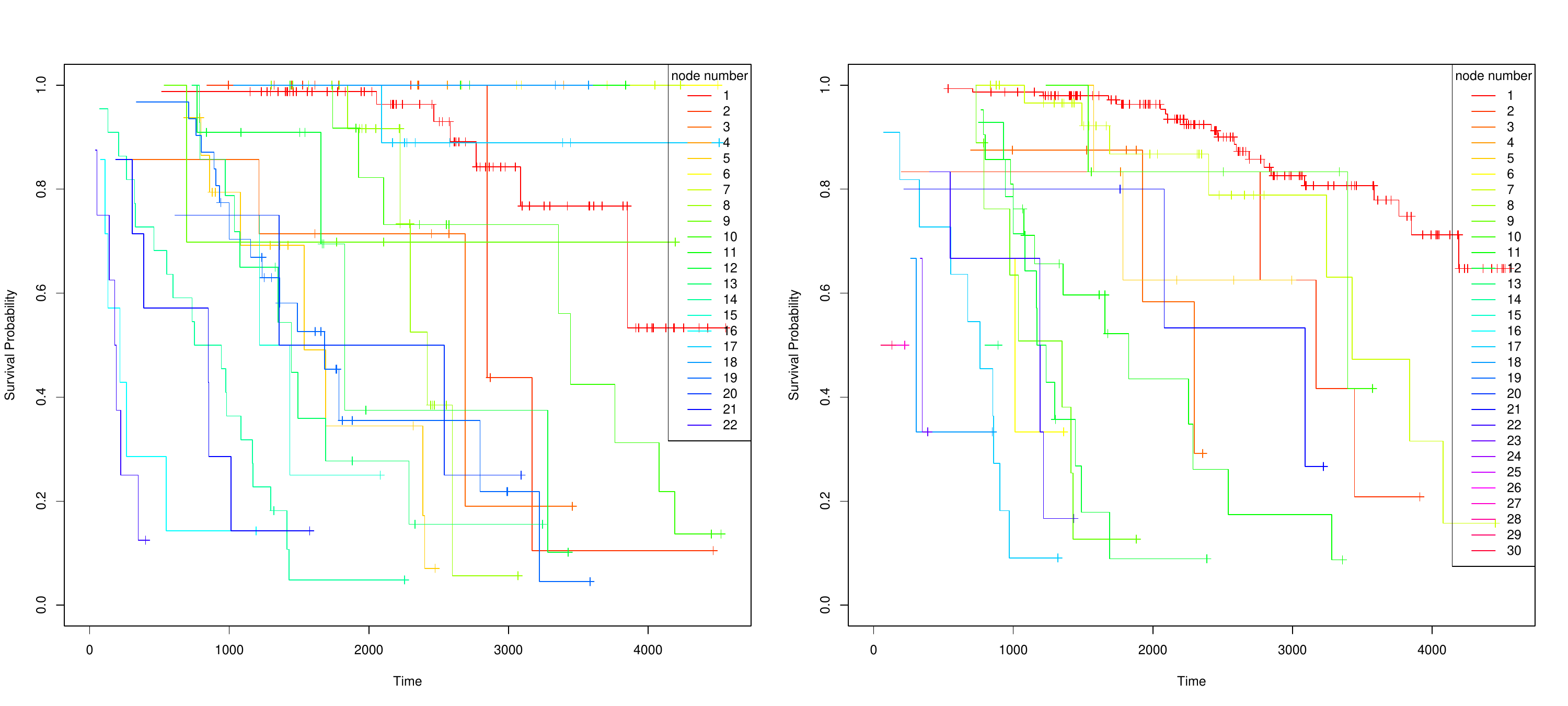}
\caption{Survival curve estimates of terminal nodes, with node 1 indicating highest survival probability. CGE tree with $\tau_{assum.}=0.375$  (left) and the logrank tree using the Kaplan-Meier estimator (right). The trees were calculated with $\tilde{p}=0.001$ .   \label{fig:terminal_node}}
\end{figure}

\begin{figure}
\centering
\includegraphics[scale=0.65]{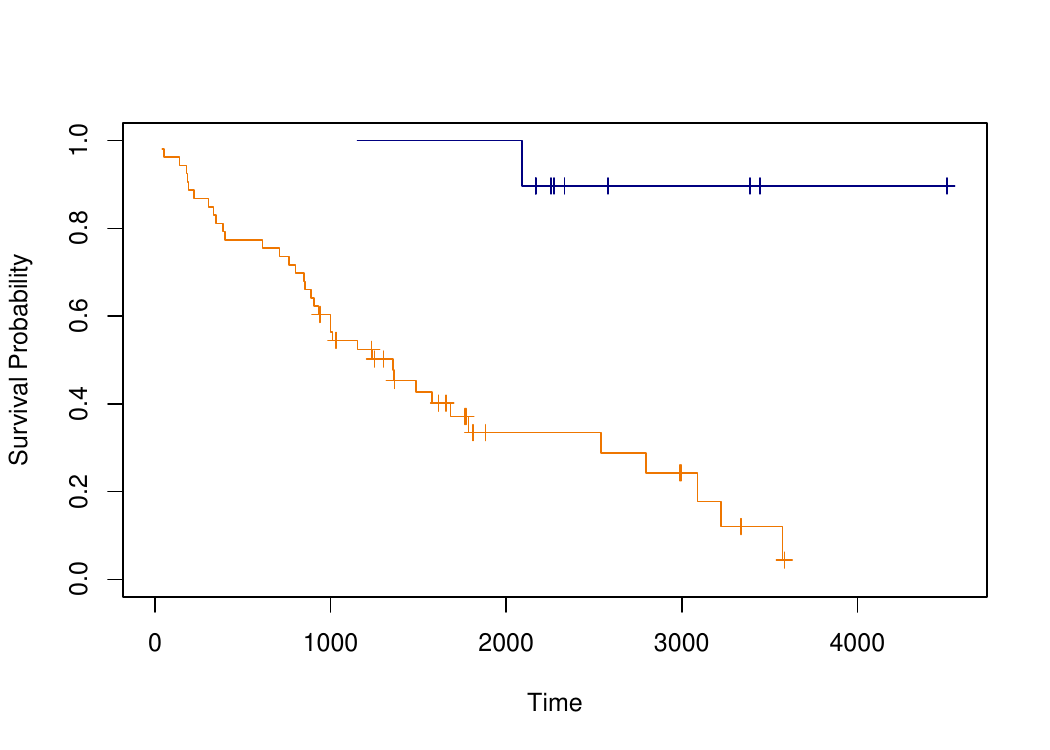}
\caption{Survival curve estimate after a second split for the CGE tree with $\tau_{assum.}=0.375$.  \label{fig:second_split}}
\end{figure}

\end{document}